\documentclass[iop]{emulateapj}

\usepackage{tabularx}
\usepackage{graphicx}
\usepackage[section]{placeins}
\usepackage{array}
\usepackage{amsmath}
\usepackage{amssymb}
\usepackage{wrapfig}
\usepackage{upgreek}
\usepackage{color}
\usepackage{booktabs}
\usepackage{verbatim}
\usepackage{bm}

\def \arcsec {$^{\prime\prime}$}

\def \um {$\upmu$m}

\def \sun {$_{\odot}$}

\def \mnras {MNRAS}
\def \aap {A\&A}
\def \aj {AJ}
\def \pasp {PASP}
\def \apj {ApJ}
\def \apjs {ApJS}
\def \apjl {ApJL}
\def \pasj {PASJ}
\def \nature {Nature}

\def \araa {ARA\&A}

\newcommand{\bfield}{6.6}
\newcommand{\bfielderr}{4.7}

\newcommand{\btot}{8.4}
\newcommand{\btoterr}{6.0}

\newcommand{\bfrootn}{7.3}
\newcommand{\bfrootnerr}{2.3}

\newcommand{\btrootn}{9.2}
\newcommand{\btrootnerr}{2.9}

\newcommand{\coldens}{3.6}
\newcommand{\coldenserr}{2.8}

\newcommand{\voldens}{0.83}
\newcommand{\voldenserr}{0.66}

\newcommand{\angdisp}{4.0}
\newcommand{\angdisperr}{0.3}

\newcommand{\displo}{0.2}
\newcommand{\disphi}{2.0}
\newcommand{\incpix}{138}

\newcommand{\magE}{4.8}
\newcommand{\magU}{1.7}

\newcommand{\mphi}{0.41}
\newcommand{\mphicorr}{0.14}

\newcommand{\alfven}{9.4}
\newcommand{\alfvenerr}{6.6}
\newcommand{\alfvendev}{4.8}
\newcommand{\alfvendeverr}{3.4}

\received{--}
\revised{--}
\accepted{--}

\shorttitle{The magnetic field strength in Orion A}
\shortauthors{Pattle et al.}

\begin{document}

\title{The JCMT BISTRO Survey: The magnetic field strength in the Orion A filament}

\author{Kate Pattle\altaffilmark{1}}
\author{Derek Ward-Thompson\altaffilmark{1}}
\author{David Berry\altaffilmark{2}}
\author{Jennifer Hatchell\altaffilmark{3}}
\author{Huei-Ru Chen\altaffilmark{4,5}}
\author{Andy Pon\altaffilmark{6}}
\author{Patrick M. Koch\altaffilmark{5}}
\author{Woojin Kwon\altaffilmark{7,8}}
\author{Jongsoo Kim\altaffilmark{7,8}}
\author{Pierre Bastien\altaffilmark{9}}
\author{Jungyeon Cho\altaffilmark{10}}
\author{Simon Coud\'{e}\altaffilmark{9}}
\author{James Di Francesco\altaffilmark{11,12}}
\author{Gary Fuller\altaffilmark{13}}
\author{Ray S. Furuya\altaffilmark{14}}
\author{Sarah F. Graves\altaffilmark{2}}
\author{Doug Johnstone\altaffilmark{11,12}}
\author{Jason Kirk\altaffilmark{1}}
\author{Jungmi Kwon\altaffilmark{15}}
\author{Chang Won Lee\altaffilmark{7,8}}
\author{Brenda C. Matthews\altaffilmark{11,12}}
\author{Joseph C. Mottram\altaffilmark{16}}
\author{Harriet Parsons\altaffilmark{2}}
\author{Sarah Sadavoy\altaffilmark{17}}
\author{Hiroko Shinnaga\altaffilmark{18}}
\author{Archana Soam\altaffilmark{7}}
\author{Tetsuo Hasegawa\altaffilmark{19}}
\author{Shih-Ping Lai\altaffilmark{4,5}}
\author{Keping Qiu\altaffilmark{20,21}}
\author{Per Friberg\altaffilmark{2}}


\altaffiltext{1}{Jeremiah Horrocks Institute, University of Central Lancashire, Preston, United Kingdom, PR1 2HE}
\altaffiltext{2}{East Asian Observatory, 660 N. A`oh\={o}k\={u} Place, University Park, Hilo, Hawaii 96720, USA}
\altaffiltext{3}{Physics and Astronomy, University of Exeter, Stocker Road, Exeter, EX4 4QL, United Kingdom}
\altaffiltext{4}{Institute of Astronomy and Department of Physics, National Tsing Hua University, Hsinchu 30013, Taiwan}
\altaffiltext{5}{Academia Sinica Institute of Astronomy and Astrophysics, P. O. Box 23-141, Taipei 10617, Taiwan}
\altaffiltext{6}{Department of Physics and Astronomy, The University of Western Ontario, 1151 Richmond Street, London, N6A 3K7, Canada}
\altaffiltext{7}{Korea Astronomy and Space Science Institute, 776 Daedeokdae-ro, Yuseong-gu, Daejeon 34055, Republic of Korea}
\altaffiltext{8}{Korea University of Science and Technology, 217 Gajang-ro, Yuseong-gu, Daejeon 34113, Republic of Korea}
\altaffiltext{9}{Centre de recherche en astrophysique du Qu\'ebec \& d\'epartement de physique, Universit\'e de Montr\'eal, C.P. 6128, Succ. Centre-ville, Montr\'eal, QC, H3C 3J7, Canada}
\altaffiltext{10}{Department of Astronomy and Space Science, Chungnam National University, 99 Daehak-ro, Yuseong-gu, Daejeon 34134, Republic of Korea}
\altaffiltext{11}{NRC Herzberg Astronomy and Astrophysics, 5071 West Saanich Rd, Victoria, BC, V9E 2E7, Canada}
\altaffiltext{12}{Department of Physics and Astronomy, University of Victoria, Victoria, BC, V8P 1A1, Canada}
\altaffiltext{13}{Jodrell Bank Centre for Astrophysics, School of Physics and Astronomy, University of Manchester, Oxford Road, Manchester, M13 9PL, United Kingdom}
\altaffiltext{14}{Institute of Liberal Arts and Sciences Tokushima University, Minami Jousanajima-machi 1-1, Tokushima 770-8502, Japan}
\altaffiltext{15}{Institute of Space and Astronautical Science, Japan Aerospace Exploration Agency, 3-1-1 Yoshinodai, Chuo-ku, Sagamihara, Kanagawa 252-5210, Japan}
\altaffiltext{16}{Max Planck Institute for Astronomy, K\"{o}nigstuhl 17, 69117 Heidelberg, Germany}
\altaffiltext{17}{Harvard-Smithsonian Center for Astrophysics, 60 Garden Street, Cambridge, MA 02138, USA}
\altaffiltext{18}{Kagoshima University, 1-21-35 Korimoto, Kagoshima, Kagoshima 890-0065, Japan}
\altaffiltext{19}{National Astronomical Observatory, National Institutes of Natural Sciences, Osawa, Mitaka, Tokyo 181-8588, Japan}
\altaffiltext{20}{School of Astronomy and Space Science, Nanjing University, 163 Xianlin Avenue, Nanjing 210023, China}
\altaffiltext{21}{Key Laboratory of Modern Astronomy and Astrophysics (Nanjing University), Ministry of Education, Nanjing 210023, China}


\label{firstpage}

\begin{abstract}
We determine the magnetic field strength in the OMC~1 region of the Orion A filament via a {new implementation of the Chandrasekhar-Fermi method} using observations performed as part of the James Clerk Maxwell Telescope (JCMT) B-Fields In Star-Forming Region Observations (BISTRO) survey with the POL-2 instrument.  {We combine BISTRO data with archival SCUBA-2 and HARP observations} to find a plane-of-sky magnetic field strength in OMC~1 of {${B_{\rm pos}=\bfield\pm\bfielderr}$\,mG, where ${\delta B_{\rm pos}=\bfielderr}$\,mG represents a predominantly systematic uncertainty}.  {We develop} a new method for measuring angular dispersion, analogous to unsharp masking.  We find a magnetic energy density of {${\sim\magU\times 10^{-7}}$\,J\,m${^{-3}}$} in OMC~1, comparable both to the {gravitational potential energy density of OMC~1} ($\sim 10^{-7}$\,J\,m$^{-3}$), and to the energy density in the Orion BN/KL outflow ($\sim 10^{-7}$\,J\,m$^{-3}$).  We find that neither the Alfv\'{e}n velocity in OMC~1 nor the velocity of the super-Alfv\'{e}nic outflow ejecta is sufficiently large for the BN/KL outflow to have caused large-scale distortion of the local magnetic field in the $\sim$500-year lifetime of the outflow.  Hence, we propose that the hour-glass field morphology in OMC~1 is caused by the distortion of a primordial cylindrically-symmetric magnetic field by the gravitational {fragmentation of the filament and/or the gravitational interaction of the BN/KL and S clumps}.  {We find that OMC~1 is currently in or near magnetically-supported equilibrium, and} that the current large-scale morphology of the BN/KL outflow is regulated by the geometry of the magnetic field in OMC~1, and not vice versa.
\end{abstract}


\section{Introduction}

The role of magnetic fields in the star formation process is currently poorly observationally constrained (e.g. \citealt{crutcher2012}).  Measurements of magnetic field strength in star-forming clouds vary from a few $\upmu$G in quiescent low-mass regions (e.g. \citealt{crutcher2000}) to $\lesssim 10$\,mG in massive molecular clouds (e.g. \citealt{curran2007}).  Nonetheless, low- and high-mass star-forming regions appear to have many commonalities in both their gas and their magnetic field morphologies.

Recent observations, particularly those made by the \emph{Herschel} Space Observatory, have shown that filaments are ubiquitous in molecular clouds (e.g. \citealt{andre2010}), and have led to the hypothesis that the dominant mode of formation of solar-mass stars is to form on dense, self-gravitating filaments \citep{andre2014}.  A recently-proposed paradigm of magnetically-regulated filamentary star formation \citep{andre2014} suggests that material flows onto filaments along magnetic field lines, until the filament has accreted sufficient mass to collapse under gravity to form a series of prestellar cores.

In low-mass star-forming regions, the magnetic field orientation {has been seen to be perpendicular to the filament direction in low-density material surrounding dense, self-gravitating, filaments.  Faint `striations' are seen} in the low-density molecular gas parallel to the magnetic field direction, suggesting that material is accreting onto filaments along magnetic field lines (\citealt{sugitani2011}; \citealt{palmeirim2013}; \citealt{matthews2014}).  Observations of high-mass star-forming regions have shown behaviour qualitatively similar to that in low-mass star-forming regions \citep{wardthompson2017}.  However, in order to accurately constrain the role of magnetic fields in high-mass filaments, and to understand the connection between the roles of magnetic fields in low- and high-mass star formation, detailed studies of the strength of magnetic fields in high-mass filaments, and their contribution to the energy balance of high-mass star-forming regions, must be undertaken.

The Orion Nebula is the nearest site of high-mass star formation to the Earth \citep{odell2008}.  The complex morphology of the region is well-resolved by modern telescopes, allowing its multiple sites of past and ongoing high-mass star formation to be studied in detail (see, e.g. \citealt{bally2008}; \citealt{odell2008}).  In this paper we are concerned with the OMC~1 region, located at a distance of $388\pm5$\,pc \citep{kounkel2017} in the centre of the `integral filament' \citep{bally1987}; a dense molecular cloud and a site of ongoing high-mass star formation.

OMC~1 is located behind the Trapezium cluster, a group of young stars containing sufficient OB stars to photoionize the surrounding gas.  The ionized gas surrounding the Trapezium cluster is bounded by the Orion Bar photon-dominated region (PDR), which we see edge-on, to the south-east of and in front of the dense gas of OMC~1 (e.g. \citealt{odell2008}).  OMC~1 consists of a large mass of submillimeter-bright dense gas, separated into two principal clumps, the northern Becklin-Neugebauer-Kleinmann-Low (BN/KL) clump (\citealt{becklin1967}; \citealt{kleinmann1967}) and the southern Orion S clump (\citealt{batrla1983}; \citealt{haschick1989}).  The BN/KL clump hosts an extremely powerful explosive molecular outflow, with a wide opening angle and multiple ejecta known as the `bullets of Orion' (\citealt{kwan1976}; \citealt{allen1993}).

The magnetic field of the OMC~1 region has an hour-glass morphology (\citealt{schleuning1998}; \citealt{houde2004}; \citealt{wardthompson2017}).  A variety of magnetic field strengths have been reported in OMC~1, ranging from a few hundred $\upmu$G (\citealt{crutcher1999}, {CN Zeeman effect, 23\arcsec\ resolution}; \citealt{houde2009}, {dust polarization, 12\arcsec\ resolution}) to a few mG ({\citealt{hansen1983}, \citealt{norris1984}, \citealt{johnston1989}, \citealt{cohen2006}, OH maser emission, 0.15-0.3\arcsec\ resolution}; \citealt{hildebrand2009}, {dust polarisation, 20\arcsec\ resolution}; \citealt{tang2010}, {energetics arguments from 1\arcsec-resolution dust continuum observations}).

In this paper we analyze observations of the OMC~1 region taken in polarized light by the POL-2 polarimeter (\citealt{friberg2016}; Bastien et al., in prep.) operating in conjunction with the SCUBA-2 (Submillimetre Common-User Bolometer Array~2) camera \citep{holland2013} on the James Clerk Maxwell Telescope (JCMT).  We use these data alongside archival JCMT photometric and spectroscopic data in order to determine the strength of the magnetic field in OMC~1 using the Chandrasekhar-Fermi method \citep{chandrasekhar1953}, and to investigate the relative importance of the magnetic field to the energy balance of OMC~1.

The POL-2 data used in this work were taken as part of the BISTRO (B-Fields in Star-Forming Region Observations) survey \citep{wardthompson2017} and as part of the POL-2 commissioning project.  The BISTRO survey is observing the high-column-density regions of the molecular clouds of the Gould Belt (\citealt{herschel1847}; \citealt{gould1879}) in polarized light, in order to produce a large and homogeneous data set for the investigation of the role of magnetic fields in the physics of star formation in nearby molecular clouds.

The structure of this paper is as follows.  In Section 2 we discuss the observations and data reduction.  In Section 3 we determine the magnetic field strength in OMC~1 using the Chandrasekhar-Fermi method.  In Section 4 we estimate the energy balance between the magnetic field, gravitational interaction, thermal and non-thermal gas motions, and outflow of OMC~1.  In Section 5 we discuss our results, and in Section 6 we summarize our conclusions.

\section{Observations}
\label{sec:observations}

\begin{figure}
\centering
\includegraphics[width=0.4\textwidth]{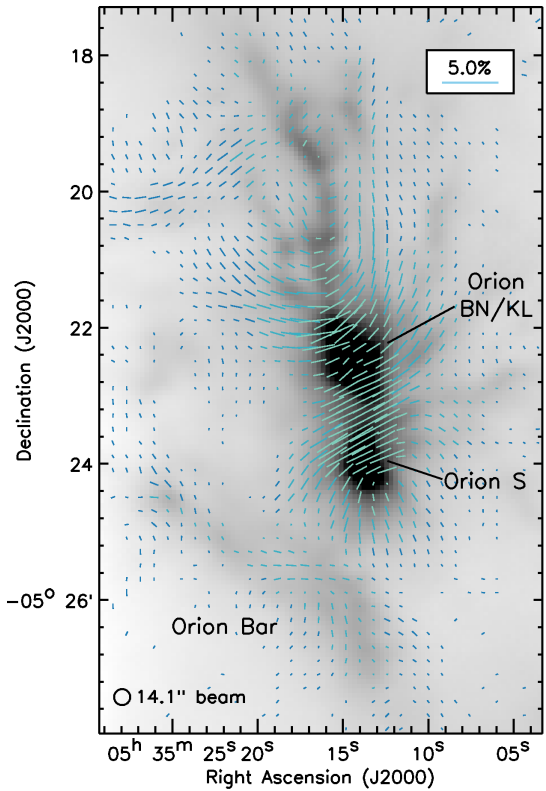}
\caption{A map of the polarization half-vectors in the centre of OMC~1, with half-vectors rotated by $90\,$degrees to show the direction of the magnetic field, modified from \citet{wardthompson2017}.  The background greyscale image is a SCUBA-2 850-\um\ total intensity image of Orion A.  The Orion BN/KL, Orion S and Orion Bar features are labelled.  Only those half-vectors with $(P/\delta P)\geq3$ are shown.  The half-vector color scale is chosen for contrast against the background image and has no physical meaning.}
\label{fig:oriona}
\end{figure}

The POL-2 observations used in this analysis were originally presented by \citet{wardthompson2017}, and form part of the JCMT BISTRO Large Program.  We refer readers to that work for a detailed description of the data reduction, and summarize the key points here.  Continuum observations in polarized light at 850\um\ were made by inserting POL-2 (Bastien et al., in prep; \citealt{friberg2016}) into the optical path of SCUBA-2 \citep{holland2013}.  The OMC~1 region was observed 21 times with POL-2 between 2016 January 11 and 2016 January 24 in a mixture of very dry weather (Grade 1; $\tau_{225\,{\rm GHz}}<0.05$) and dry weather (Grade 2; $0.05\leq\tau_{225\,{\rm GHz}}<0.08$), providing a total of 14 hours of on-source integration.  The JCMT has an effective beam size of 14.1 arcsec at 850\um, equivalent to 0.027\,pc at a distance of 388\,pc.

The 850\um\ data were reduced in a two-stage process.  The raw bolometer timestreams were first converted to separate Stokes $Q$ and Stokes $U$ timestreams using the process \emph{calcqu} in \textsc{smurf} \citep{berry2005}.  The $Q$ and $U$ timestreams were then reduced separately using an iterative map-making technique, \emph{makemap} in \textsc{smurf} \citep{chapin2013} and gridded to 4-arcsec pixels.  The iterations were halted when the map pixels, on average, changed by $<\,5$ per\,cent of the estimated map RMS noise.  In order to correct for the instrumental polarization (IP), \emph{makemap} is supplied with a total intensity image {(${I}$)} of the source, taken using SCUBA-2 while POL-2 is not in the beam (Bastien et al. in prep.; \citealt{friberg2016}).  We took our total intensity image of OMC~1 from a SCUBA-2 observation made using the standard SCUBA-2 DAISY mapping mode.

The Stokes $Q$ and $U$ observations were combined using the process \emph{pol2stack} in \textsc{smurf} to produce an output half-vector catalogue (`half-vector' refers to the $\pm180$ degree ambiguity in magnetic field direction).  The half-vectors which we use in this work are gridded to a 12-arcsec pixel size to improve signal-to-noise.  Throughout this work we use polarization half-vectors rotated by $90\,$degrees to trace the magnetic field direction, hereafter referred to as `magnetic field half-vectors'.

The absolute calibration of the data is discussed by \citet{wardthompson2017}.  In this work we use the measured magnetic field angles, {${\theta = 0.5\arctan(U/Q)}$}, and polarization fraction, ${P = (Q^{2}+U^{2}-0.5[(\delta Q)^{2}+(\delta U)^{2}])^{0.5}/I}$, in OMC~1. {${P}$ is debiased using the mean of the ${Q}$ and ${U}$ variances, ${(\delta Q)^{2}}$ and ${(\delta U)^{2}}$ respectively.  We note that there are many methods for debiasing polarization data (see, e.g. Montier 2015a,b\nocite{montier2015a}\nocite{montier2015b}).  However, in this work we use $P$ for half-vector selection only, so the effect of our choice of debiasing method on our results is minimal.}  The measured magnetic field angles are determined from the relative values of the Stokes $Q$ and $U$ parameters, and hence do not depend on the absolute calibration (i.e. the polarized intensity) of the data.  

\section{Results}

\begin{figure*}
\centering
\includegraphics[width=0.8\textwidth]{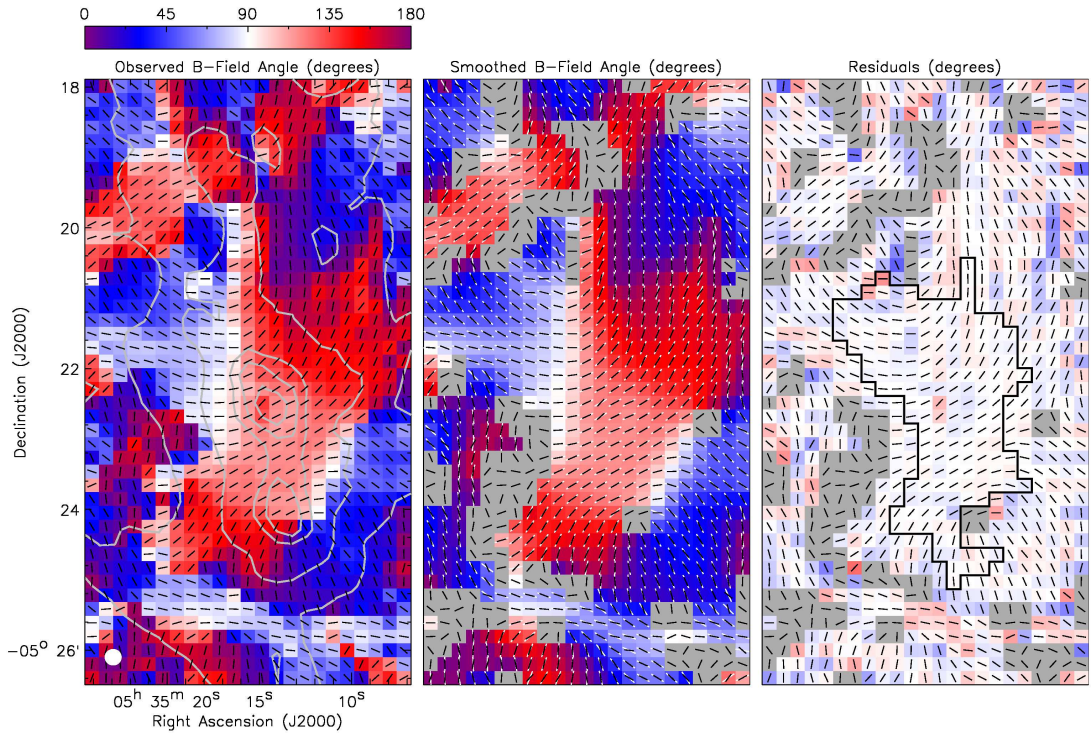}
\caption{Position angles of magnetic field half-vectors in OMC~1.  The color maps in each panel show as follows: left-hand panel -- the observed magnetic field position angles, $\theta_{\rm obs}$; central panel -- {mean} magnetic field position angle, $\langle\theta\rangle$, from smoothing $\theta_{\rm obs}$ with a $3\times 3$-pixel boxcar filter; right-hand panel -- the residual position angle values, $\theta_{\rm obs}-\langle\theta\rangle$.  The colour scale is the same in all panels, and is chosen to wrap around the discontinuity at $0/180$ degrees.  In all panels the measured magnetic field half-vectors (as shown in Figure~\ref{fig:oriona}) are marked in black.  In the central panel, the smoothed magnetic field position angles are marked in white, for comparison.  In the central and right-hand panels, pixels excluded due to changes of angle $\geq 90$ degrees in their boxcar window are marked in grey.  The region outlined in black in the right-hand panel marks the pixels over which the magnetic field strength in OMC~1 is measured (i.e. where $\delta(\Delta\theta)<1.8$ degrees, as discussed in the text).  Stokes ${{I}}$ emission is shown as grey contours on the left-hand panel.  The JCMT 850-\um\ beam is marked as a white circle in the lower left-hand corner of the left-hand panel.}
\label{fig:angles}
\end{figure*}

We determined the magnetic field strength in OMC~1 using the Chandrasekhar-Fermi (CF; \citealt{chandrasekhar1953}) method.  The CF method assumes that the underlying magnetic field geometry is uniform, and that the dispersion of measured polarization angles (after any necessary correction for measurement errors) represents the distortion of the magnetic field by turbulent and other motions in the gas.

We determined the plane-of-sky magnetic field strength ($B_{\rm pos}$) in OMC~1 using the formulation of the CF method given by \citet{crutcher2004}:
\begin{equation}
B_{\rm pos} = Q^{\prime}\,\sqrt{4\pi\rho}\,\frac{\sigma_{v}}{\sigma_{\theta}}\approx 9.3\,\sqrt{n({\rm H}_{2})}\,\frac{\Delta v}{\langle\sigma_{\theta}\rangle}\,\upmu{\rm G},
\label{eq:cf}
\end{equation}
where $\sigma_{v}$ is the one-dimensional non-thermal velocity dispersion in the gas; $\sigma_{\theta}$ is the dispersion in polarization position angles; $\rho$ is the gas density; $\Delta v$ is the FWHM velocity dispersion in ${\rm km}\,{\rm s}^{-1}$ ($\Delta v=\sigma_{v}\sqrt{8{\rm ln}(2)}$); $\langle\sigma_{\theta}\rangle$ is the typical deviation in polarization position angle in degrees; $n({\rm H}_{2})$ is the number density of molecular hydrogen ($\rho=\mu m_{\textsc{h}}n({\rm H}_{2})$, where $\mu$ is the mean molecular weight of the gas); and {${Q^{\prime}}$} is a factor of order unity accounting for variation in field strength on scales smaller than the beam {(labelled ${Q^{\prime}}$ to distinguish it from the Stokes ${Q}$ parameter)}.  \citet{crutcher2004} take ${Q}^{\prime}=0.5$ (c.f. \citealt{ostriker2001}).  We adopt this value throughout this paper.  We discuss the appropriate value of the $Q^{\prime}$ parameter in Section~\ref{sec:q}, below.

\citet{crutcher2004} note that the CF method does not constrain the line-of-sight component of the magnetic field strength, and that statistically,
\begin{equation}
B_{\rm pos}\approx\frac{\pi}{4}|\mathbf{B}|
\label{eq:b_correction}
\end{equation}
on average, where $|\mathbf{B}|$ is the magnitude of the magnetic field strength half-vector.  However, this statistical correction assumes that the magnetic field has a large-scale geometry that is not biased by a preferred axis.  The magnetic field in Orion A is clearly highly ordered (see Figure~\ref{fig:oriona}), and so we cannot rule out a preferred orientation for the line of sight field.  The relevance of this correction to the plane-of-sky field strength that we measure is hence unclear.  We discuss this further below.

\subsection{Angular dispersion in OMC~1} 
\label{sec:angdisp}

\begin{figure}
  \centering
  \includegraphics[width=0.4\textwidth]{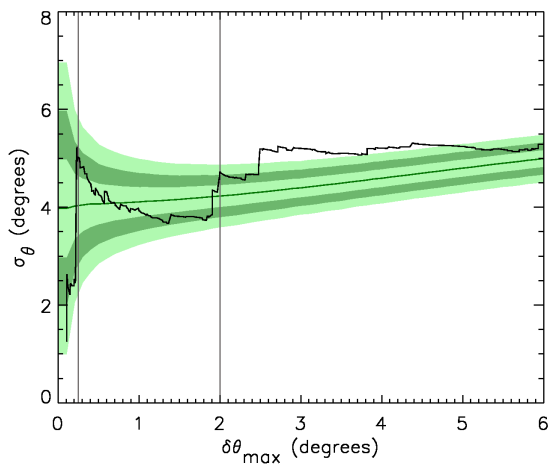}
  \caption{The standard deviation, $\sigma_{\theta}$, of the distribution of the magnetic field angles in the OMC~1 data about the mean magnetic field direction as a cumulative function of maximum allowed uncertainty in the $3\times3$-pixel smoothing box, $\delta\theta_{max}$ (black), plotted on a Monte Carlo simulation of the measured standard deviation distribution for a data set with a true standard deviation of {4.0} degrees (dark green).  Green shaded regions show the 1-, 2-, and 3-$\sigma$ uncertainties on the Monte Carlo simulation.  Vertical grey lines show the region over which the average angular dispersion $\langle\sigma_{\theta}\rangle$ is measured.}
  \label{fig:ang_dev}
\end{figure}

In order to apply the CF method to the magnetic field in OMC~1, which is both highly ordered and significantly non-uniform, it is necessary to remove or account for the effect of the underlying field geometry before estimating the dispersion in position angle.  We present a method for measuring angular dispersion in an ordered field which is analogous to unsharp masking: we estimated the behaviour of the non-distorted magnetic field by applying a smoothing function to our polarization angle map.  We then subtracted our {estimated} non-distorted {(i.e. smoothed)} magnetic field directions from the measured polarization angles (rotated by 90 degrees to trace magnetic field direction) in order to find the difference between the measured magnetic field angle and the mean field direction in each pixel in the map.

We subtract the smoothed map ($\langle\theta\rangle$) from the map of measured position angle ($\theta_{obs}$), giving a residual map showing the deviation in angle in each pixel from the mean field direction, $\Delta\theta$, i.e.
\begin{equation}
\Delta\theta = \theta_{obs} - \langle\theta\rangle.
\end{equation}
The observed and smoothed position angle maps, and their residual, are shown in Figure~\ref{fig:angles}.

We estimate mean field directions by smoothing the map of measured angles using a $3\times3$-pixel boxcar average.  The $3\times 3$-pixel boxcar filter was chosen in order to allow a smoothing length smaller than the radius of curvature of the magnetic field in the high-signal-to-noise regions of Orion A.  We measure polarization angles in the range  $0<\theta\leq 180$ degrees, measuring angles east of north.

The 180-degree ambiguity in magnetic field direction, which is inherent in polarimetric observations, introduces a discontinuity in the distribution of angles.  For our choice of range of angles, this discontinuity occurs at 0 or 180 degrees.  In order to avoid creating artefacts in our smoothed map due to averaging over groups of pixels within which this discontinuity is crossed, we tested each $3\times 3$-pixel boxcar in order to determine whether the greatest difference in angle between pixels within the boxcar was $\geq 90$ degrees.  If this was the case, we mapped the pixels within that boxcar from the range $0<\theta\leq 180$ to the range $-90<\theta\leq 90$ degrees, and repeated the test of maximum difference in angle.  If the maximum difference in angle remained $\geq 90$ degrees after this mapping, then we concluded that the observed variation in angle was real and that the field in the vicinity of that pixel was insufficiently uniform over the boxcar for the smoothing function to be valid, and so excluded that pixel from further analysis.  However, if the mapping reduced the maximum difference in angle to a range $< 90$ degrees, then we treated that boxcar as containing pixels which cross the $0/180$-degree discontinuity, and determined the average position angle from the angles mapped to the range $-90<\theta\leq 90$ degrees.  Where necessary, we then reversed the mapping in angle.

 The pixels which we exclude from the analysis are marked in grey in the central and right-hand panels of Figure~\ref{fig:angles}.  These pixels represent a small fraction of the total number of well-characterised pixels in OMC~1.

 We investigated how the measured standard deviation of the distribution of deviation angles, $\sigma_{\theta}$, varies as a function of uncertainty in deviation angle $\delta(\Delta\theta)$ for the general case of a Gaussian distribution of angles each with and associated experimental uncertainty by performing Monte Carlo simulations of data sets with a range of fixed underlying dispersions and randomly generated measurement errors.  We found that, when measured over well-characterized pixels, the measured standard deviation tends closely to the true underlying standard deviation.  If measuring over poorly-characterized pixels, the measured standard deviation increases linearly with maximum allowed uncertainty on angle.  These results are shown in Appendix~A.

We tested the validity of our `unsharp masking' method of recovering angular dispersion by testing it on sets of synthetic observations with various field curvatures, intrinsic angular dispersions, and measurement uncertainties.  The results of these tests are shown in Appendix~B.  We find that our `unsharp masking' method accurately recovers the true angular dispersion of the data provided that the systematic variation in field direction over the box size due to the changing direction of the underlying field is significantly smaller than the random variation in field direction due to the dispersion on position angle.  We find that we are in this regime throughout the central region of OMC~1, and so the measured angular dispersion should be an accurate estimate of the intrinsic angular dispersion in the data.

We tested the effect of measurement uncertainties on our recovery of angular dispersion using the unsharp-masking method, and found that as in the generalised case, ${\sigma_{\theta}}$ is not altered by measurement errors provided that those measurement errors are small.  However, as previously, the measured angular dispersion increases approximately linearly with measurement uncertainty if the measurement uncertainty is comparable to or greater than the angular dispersion.  This is true regardless of the degree of field curvature.

We find that for angular dispersions of $\sim 4$ degrees, the effect of measurement error on $\sigma_{\theta}$ is minimal while $\delta\theta_{max}\lesssim 2$ degrees, where $\delta\theta_{max}$ is the maximum uncertainty in any pixel included in the smoothing box.  We thus restrict our application of the unsharp-masking method in OMC~1 to those pixels for which $\delta\theta_{max}< 2.0$ degrees.  Uncertainties on position angle are calculated by \textit{pol2stack} from the variances on the $Q$ and $U$ values in each pixel in the coadded $Q$ and $U$ maps from which the vector properties are calculated, using standard error propagation (see Section~\ref{sec:observations}).  We are therefore confident that our method is valid in this case.

Taking the mean of the standard deviations of the distributions containing only the best-characterized pixels ($\displo<\delta\theta_{max}<\disphi$ degrees; up to \incpix\ pixels), we find a mean dispersion of $\langle\sigma_{\theta}\rangle = \angdisp\pm \angdisperr$ degrees.  The measured angular dispersion $\sigma_{\theta}$ is plotted as a cumulative function of $\delta\theta_{max}$ in Figure~\ref{fig:ang_dev}.

The pixels with low measurement uncertainties form a contiguous region with low residuals, marked in the right-hand panel of Figure~\ref{fig:angles}.  The variation in $\delta(\Delta\theta)$ across OMC~1 is shown explicitly in Figure~\ref{fig:figa6} in Appendix~A.  This contiguous region includes the high-density region of OMC~1: the BN/KL and S regions, the space between them, and much of the region in which the magnetic field shows an hour-glass morphology.

  While there is variation in the dispersion of magnetic field half-vectors about the mean field direction across OMC~1, our data are sufficiently well-characterized that we have a statistically-significant sample of good measurements in the centre of the OMC~1 molecular cloud, the region of most interest for our scientific analysis.  We thus adopt the angular dispersion which we consistently measure across this region, $\langle\sigma_{\theta}\rangle = \angdisp\pm\angdisperr$ degrees, throughout the rest of this study.  We henceforth restrict our analysis to the dense centre of the OMC~1 cloud, containing the BN/KL and S clumps.

\subsection{Velocity dispersion in OMC~1}
\label{sec:vel_disp}

We determined the average velocity dispersion in the gas in OMC~1 from the HARP (Heterodyne Receiver Array Program; \citealt{buckle2009}) C$^{18}$O {${J=3\to2}$} measurements of OMC~1 presented by \citet{buckle2012}.  {HARP is mounted on the JCMT, and hence the C$^{18}$O observations, with a rest frequency of 329.33\,GHz \citep{cdms}, have the same resolution as the POL-2 850-\um\ data.}  We assume that C$^{18}$O traces approximately the same material as the 850-\um\ dust emission.  As C$^{18}$O traces number densities up to a few times $10^{5}$\,cm$^{-3}$ (e.g. \citealt{difrancesco2007}), comparable to the median value we determine in OMC~1 (see below), this assumption should be valid.  We fitted the C$^{18}$O data in the manner described by \citet{pattle2015}: we fitted a single Gaussian to each pixel, accepting fits with a signal-to-noise ratio $\ge 5$.  We took the Gaussian width of the fit to be the 1D velocity dispersion in C$^{18}$O in that pixel.

{The C$^{18}$O data in OMC~1 are generally well-fitted by a single Gaussian, particularly on the bright central filament where the majority of the mass lies.  There are a few positions at which the C$^{18}$O data show double peaks or broad wings suggestive of outflow contamination, but these are typically found off-filament in low-density and low-signal-to-noise regions which are also coincident with the BN/KL outflow (discussed below).  These regions are generally excluded from our fitting by the S/N cut we apply, and we expect outflow contamination to have minimal effect on the mean velocity dispersion that we measure over the region.}

We converted the C$^{18}$O velocity dispersions to {non-thermal} velocity dispersions using the relation
\begin{equation}
{\sigma_{v}^{2}=\sigma_{v,\textsc{c}^{18}\textsc{o}}^{2}-\frac{k_{\textsc{b}}T}{m_{\textsc{c}^{18}{\textsc{o}}}},}
\end{equation}
where $\sigma_{v}$ is the {non-thermal} gas velocity dispersion, $\sigma_{v,\textsc{c}^{18}\textsc{o}}$ is the velocity dispersion of C$^{18}$O, $m_{\textsc{c}^{18}{\textsc{o}}}$ is the mass of the C$^{18}$O molecule ($m_{\textsc{c}^{18}{\textsc{o}}}=30$\,amu), and all other symbols are as defined previously.  The temperature in each pixel was taken to be the temperature at that position determined using equation~\ref{eq:temperature}, below, {(Typical temperature values are found to be $\sim 15$\,K, as discussed below.)}

We measured a mean 1D {C$^{18}$O} velocity dispersion of $\sigma_{v,\textsc{c}^{18}\textsc{o}} = 1.33\pm 0.31$\,km\,s$^{-1}$, {and a mean 1D non-thermal gas velocity dispersion of ${\sigma_{v} = 1.32\pm 0.31}$\,km\,s$^{-1}$  (${\Delta v=3.12\pm0.73}$\,km\,s$^{-1}$)} over the area we defined above, where the uncertainty is the standard deviation on the mean.  This is very similar to the mean 1D gas velocity dispersion of 1.24\,km\,s$^{-1}$ determined across the integral filament by \citet{buckle2012}.  {The gas in OMC~1 is highly supersonic, and so the contribution of thermal motions to the total linewidth is minimal.}

\subsection{Volume density of OMC~1}
\label{sec:vol_density}

We determined the average number density of particles in the OMC~1 region using SCUBA-2 450-\um\ and 850-\um\ observations presented by \citet{mairs2016}, which were taken as part of the SCUBA-2 Gould Belt Survey \citep{wardthompson2007}.  We determined column densities by repeating the method described by \citet{salji2015}, using the OMC~1 maps presented by \citet{mairs2016}.  {We chose to determine column densities from SCUBA-2 data in order to perform our analysis as self-consistently as possible.}  Contamination of the measured SCUBA-2 850-\um\ flux density by emission from the $^{12}$CO $J=3\to2$ transition can reach fractions $\sim 15-20$\% in OMC~1 \citep{coude2016}, and so we used an 850-\um\ map which has been corrected for CO contamination in the manner described by \citet{sadavoy2013}.  Before performing the following analysis we convolved the 450-\um\ SCUBA-2 map to the 850-\um\ resolution using a convolution kernel based on the model JCMT beams as described by \citet{pattle2015}, using the method introduced by \citet{aniano2011}.

We assumed that the dust in OMC~1 is optically thin and emits as a modified blackbody,
\begin{equation}
{I_{\nu}= \mu m_{\textsc{h}}\kappa(\nu)\,N({\rm H}_{2})\,B_{\nu}(T),}
\label{eq:intensity}
\end{equation}
where $I_{\nu}$ is the intensity at frequency $\nu$, $\mu=2.86$ is the mean molecular weight per hydrogen molecule, assuming that the gas is $\sim 70$\% hydrogen by mass (c.f. \citealt{kirk2013}), $m_{\textsc{h}}$ is the mass of a hydrogen atom, $N({\rm H}_{2})$ is the column density of molecular hydrogen, $B_{\nu}(T)$ is the Planck function at dust temperature $T$, and $\kappa(\nu)$ is the dust mass opacity function \citep{hildebrand1983}. $\kappa(\nu)$ is then given by
\begin{equation}
\kappa(\nu)=\kappa_{\nu_{0}}\,\left(\frac{\nu}{\nu_{0}}\right)^{\beta},
\end{equation}
where $\kappa_{\nu_{0}}$ is the dust opacity at the reference frequency $\nu_{0}$ and $\beta$ is the dust emissivity index.  We take $\kappa_{\nu_{0}}=0.1$\,cm$^{2}$\,g$^{-1}$ at $\nu_{0}=1$\,THz, assuming a dust-to-gas ratio of 1:100 \citep{beckwith1991}, and take $\beta=2.0$ {\citep{draine1984}}.

We determined a temperature for each pixel from the ratio of 850-\um\ flux density ($I_{850}$) to 450-\um\ flux density ($I_{450}$) using the implicit relation
\begin{equation}
\frac{I_{850}}{I_{450}}=\left(\frac{\nu_{850}}{\nu_{450}}\right)^{3+\beta}\times\frac{e^{\frac{h\nu_{450}}{k_{\textsc{b}}T}}-1}{e^{\frac{h\nu_{850}}{k_{\textsc{b}}T}}-1},
\label{eq:temperature}
\end{equation}
which we solved using a look-up table for each pixel in the map.  We then solved equation~\ref{eq:intensity} for column density, using the temperatures we estimated using equation~\ref{eq:temperature}.

We excluded all pixels for which $T>50\,$K, as in these cases both the 450-\um\ and 850-\um\ data points would tend toward the Rayleigh-Jeans tail of the blackbody function, and so equation~\ref{eq:temperature} would be insensitive to temperature.

We defined a rectangular area for OMC~1 centred on R.A. $05^{h}35^{m}15^{s}$ Dec. $-5^{\circ}23^{\prime}05^{\prime\prime}.75$ with angular width $1^{\prime}38^{\prime\prime}$ and angular height $3^{\prime}4^{\prime\prime}.5$, corresponding to 0.18\,pc and 0.35\,pc respectively at a distance of 388\,pc.  We measured the median H$_{2}$ column density in this area to be {${\coldens\pm\coldenserr\times10^{23}}$}\,cm$^{-2}$.  Column density varies by several orders of magnitude across OMC~1, in the range $10^{22}-10^{25}\,$cm$^{-3}$.  Hence, we consider our median column density value to be representative of typical conditions in OMC~1.

The uncertainty on our column density measurement is dominated by systematic uncertainties on the dust emission model.  We estimated the uncertainty on our column density by conservatively assuming that the reference dust opacity ${\kappa_{\nu_{0}}}$ is accurate to ${\sim 50}$\,\% (e.g. \citealt{roy2014}), that the dust opacity index ${\beta}$ has an uncertainty of approximately ${\pm 0.3}$, representative of the range of dense-gas ${\beta}$ values common in the literature (see, e.g., \citealt{schnee2010}; \citealt{planck2011_pgcc}; \citealt{sadavoy2016}), and that the uncertainty on the 850-\um\ and 450-\um\ flux densities are dominated by their calibration uncertainties, of 5\,\% and 10\,\% respectively \citep{dempsey2013}.  Propagating these uncertainties through equations~\ref{eq:intensity}--\ref{eq:temperature}, we found a median fractional systematic uncertainty in column density of 79\,\% over our defined area in OMC~1.  Thus, we take our median column density to be ${(\coldens\pm\coldenserr)\times 10^{23}}$\,cm$^{-2}$.

We assume that OMC~1 is a cylindrical filament with radius $r=0.09$\,pc and length $L=0.35$\,pc, and hence volume $\pi r^{2}L$, and that the area which we defined is the projection of that volume onto the plane of the sky, with area $2rL$.  The volume density of the filament is then related to the median column density by
\begin{equation}
n({\rm H}_{2})=\frac{2N({\rm H}_{2})}{\pi r}\cos i,
\end{equation}
where $i$ is the inclination angle of the filament to the plane of the sky.  We assume that the filament is close to the plane of the sky, i.e. $\cos i\approx 1$.  The plane-of-sky morphology of OMC~1 does not suggest that the filament is significantly elongated along the line of sight.  However, we note that if the filament were inclined at 45 degrees to the plane of the sky, the volume density would decrease by a factor of $\sqrt{2}$, and the inferred magnetic field strength would decrease by a factor of {1.19}.

For our median column density value of {${N({\rm H}_{2})=(\coldens\pm \coldenserr)\times10^{23}}\,$cm$^{-2}$}, we determined a representative volume density in OMC~1 of {${n({\rm H}_{2})\approx(\voldens\pm \voldenserr)\times10^{6}}$\,cm$^{-3}$}.  {If our assumed cylindrical geometry is correct, then the uncertainty on our estimate of column density will also be relevant to our estimate of volume density.}

\subsection{Magnetic field strength in OMC~1}

\begin{table}
\centering
\caption{Measured properties relevant to the CF analysis in OMC~1.}
\begin{tabular}{ccc}
\toprule
 Property & Symbol & Value \\
\midrule
Angular dispersion & $\langle\sigma_{\theta}\rangle$ & {${\angdisp\pm\angdisperr}$} degrees\\
FWHM velocity dispersion  & $\Delta v$ & $3.12\pm0.73$\,km\,s$^{-1}$ \\
Hydrogen column density & $N({\rm H}_{2})$ & {${\coldens\pm\coldenserr\times 10^{23}}$}\,cm$^{-2}$ \\
Hydrogen volume density & $n({\rm H}_{2})$ & {${\voldens\pm\voldenserr\times 10^{6}}$}\,cm$^{-3}$ \\
POS magnetic field strength & $B_{\rm pos}$ & {${\bfield\pm \bfielderr}$}\,mG \\ 
\bottomrule
\end{tabular}
\label{tab:magfield}
\end{table}

Using equation~\ref{eq:cf} with our measured values of $\Delta v=3.12\pm 0.73$\,km\,s$^{-1}$ and $\langle\sigma_{\theta}\rangle=\angdisp\pm\angdisperr$ degrees, we determined the relationship between plane-of sky magnetic field strength and gas volume density to be
\begin{equation}
\frac{B_{\rm pos}}{\sqrt{n}} = 9.3\frac{\Delta v \,\,({\rm km\,s^{-1}})}{\langle\sigma_{\theta}\rangle \,\, ({\rm degrees})} \upmu{\rm G\,cm}^{-\frac{3}{2}} = {{\bfrootn\pm\bfrootnerr\,\, \upmu{\rm G\,cm}^{-\frac{3}{2}}}},
\end{equation}
and between total magnetic field strength and gas volume density to be
\begin{equation}
\frac{|\mathbf{B}|}{\sqrt{n}} =\frac{4}{\pi} \frac{B_{\rm pos}}{\sqrt{n}} = {{\btrootn\pm\btrootnerr\,\, \upmu{\rm G\,cm}^{-\frac{3}{2}}}}.
\end{equation}

For our representative gas density in OMC~1, {${n({\rm H}_{2})=\voldens\pm\voldenserr\times 10^{6}}$}\,cm$^{-3}$, we determined the plane-of-sky magnetic field strength in the OMC~1 region to be {${B_{\rm pos}=\bfield\pm\bfielderr}$}\,mG.

The stated uncertainty on ${B_{pos}}$ was determined by combining the uncertainties on ${n({\rm H}_{2})}$, ${\Delta v}$ and ${\langle\sigma_{\theta}\rangle}$ given above using the standard total-derivative method of error propagation, rather than adding the fractional uncertainties in quadrature (as is sometimes done when multiplying a set of values with associated statistical uncertainties).  This conservative method was chosen in order to demonstrate the full range of ${B_{pos}}$ values which are consistent with our measurements.  The uncertainty is on ${B_{pos}}$ is dominated by the systematic uncertainty on ${n({\rm H}_{2})}$, and so our uncertainty ${\delta B_{pos}=\bfielderr}$\,mG is likewise predominantly systematic, representing an absolute range ${B_{pos}=1.9 - 11.3}$\,mG in OMC~1, rather than a 1-${\sigma}$ statistical uncertainty.  Throughout this analysis we have attempted to treat our uncertainties robustly.  We emphasize that no other analysis of a similar type ever published will be free of (frequently unacknowledged) uncertainties of this order of magnitude.  We can state that our results suggest a field strength in OMC~1 of a few mG with sufficient certainty to allow us to perform an order-of-magnitude energetics analysis of the region.  We proceed taking ${B_{pos}=\bfield}$\,mG to be representative of the magnetic field strength in OMC~1.

If equation~\ref{eq:b_correction} is relevant to Orion, then we can infer a typical total magnetic field strength in OMC~1 of {${|\mathbf{B}|=\btot\pm\btoterr}$}\,mG.  However, as the line-of-sight geometry of the magnetic field is not known, we consider the plane-of-sky field strength only for the remainder of this work, noting that the total magnetic field strength is likely to be of the same order of magnitude, and that the correction to the magnetic field strength described by equation~\ref{eq:b_correction} would not alter our conclusions.

The magnetic field half-vectors in OMC~1 are clearly highly ordered, suggesting that the magnetic field contributes significantly to the energy balance in OMC~1.  We discuss this further below.  We summarize the values used in the CF magnetic field strength calculation in Table~\ref{tab:magfield}, for reference.

\section{Energetics Calculations} 

\begin{figure}
\centering
\includegraphics[width=0.4\textwidth]{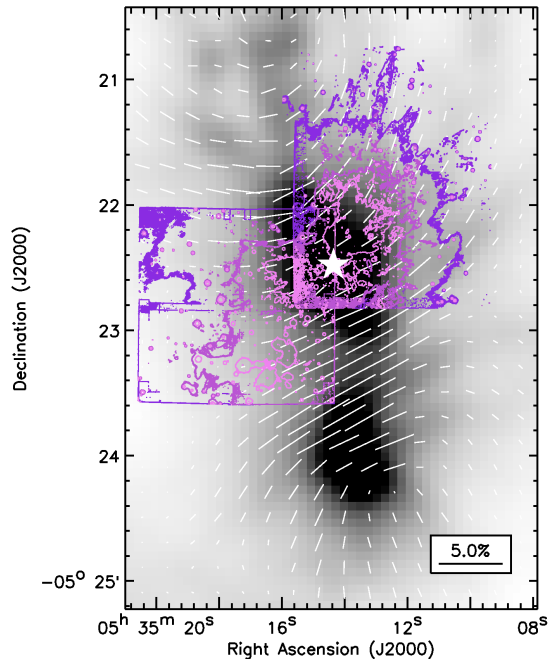}
\caption{The polarization half-vectors, rotated by 90 degrees to show magnetic field direction, (white) of OMC~1 and contours of H$_{2}$ emission \citep{bally2015} showing the BN/KL outflow, overlaid on SCUBA-2 850-\um\ emission.  The white star marks the position of the centre of the BN/KL outflow \citep{bally2015}.}
\label{fig:outflow}
\end{figure}

We infer a very strong magnetic field in the OMC~1 region, as discussed in the section above.  However, the hour-glass field morphology shown in Figure~\ref{fig:oriona} suggests that the magnetic field does not dominate the energy budget of OMC~1, as it appears to show significant deviation from the cylindrical magnetic field geometry that has previously been seen in dense filaments (\citealt{palmeirim2013}; \citealt{matthews2014}).

Other sources of energy in the OMC~1 region include gravitational potential energy, particularly that of the Orion BN/KL and Orion S clumps (the northern and southern bright regions in Figure~\ref{fig:oriona}, respectively), and energy injected by the BN/KL outflow (shown in Figure~\ref{fig:outflow}).

\begin{figure*}
\centering
\includegraphics[width=0.7\textwidth]{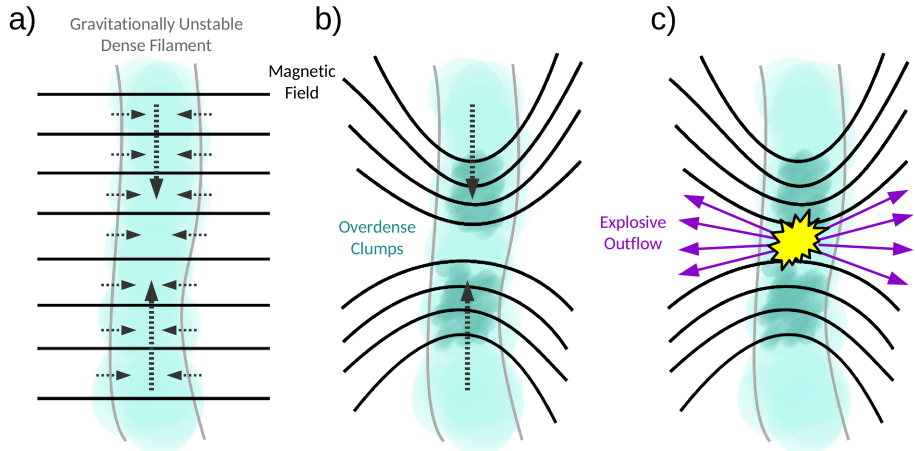}
\caption{{A cartoon of one of our proposed formation mechanisms for the hour-glass magnetic field morphology of OMC~1 and the orientation of the BN/KL outflow, in which the magnetic field is shaped by the gravitational collapse of the Orion A filament.  The magnetic field is initially cylindrically-symmetric and is frozen into the gas.  The gravitationally-unstable filament collapses axisymmetrically, dragging the flux-frozen magnetic field to create an hourglass-shaped pinch.}}
\label{fig:cartoon_gravity1} 
\end{figure*}

\begin{figure*}
\centering
\includegraphics[width=0.7\textwidth]{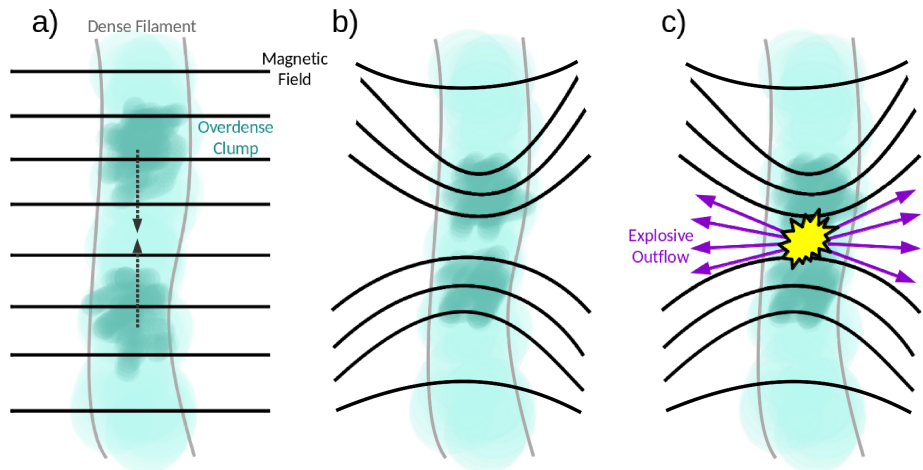}
\caption{A cartoon of one of our proposed formation mechanisms for the hour-glass magnetic field morphology of OMC~1 and the orientation of the BN/KL outflow, in which the magnetic field is shaped by the gravitational interaction of Orion BN/KL and S.  The magnetic field is initially cylindrically-symmetric and is frozen into the gas.  The two clumps are gravitationally attracted towards each other, and drag the magnetic field frozen into them along with them, while leaving the field in the lower-density filament largely undeviated.}
\label{fig:cartoon_gravity} 
\end{figure*}

If the energy budget in OMC~1 were dominated by the gravitational potential energy of the BN/KL and S clumps, the field geometry might be caused by some combination of axisymmetric collapse of the Orion A filament and of the two clumps moving toward each other, both of which mechanisms would result in the field being dragged from an initially cylindrically-symmetric morphology into the hour-glass morphology seen.  These formation mechanisms are illustrated in Figures~\ref{fig:cartoon_gravity1} and \ref{fig:cartoon_gravity}.  Similar movement of material along filaments has been observed and inferred from a combination of spectroscopic data and simulations (e.g. \citealt{balsara2001}).  Measurements of the line-of-sight velocity of the filament in isotopologues of CO \citep{buckle2012} do not rule out large-scale motion of material along the filament: the S clump appears to be moving towards us relative to the filament, while BN/KL shows no motion relative to the filament.  These two gravity-mediated formation mechanisms for the hourglass field are distinct: in the former, the BN/KL clumps and the hourglass form contemporaneously from a gravitationally unstable filament, while in the latter, the hourglass forms as a result of the gravitational interaction of the pre-existing clumps.  However, the overall effect of each mechanism on the observed magnetic field morphology is qualitatively very similar, and present-day observations cannot distinguish between these two histories.

If the energy budget in OMC~1 were dominated by the BN/KL outflow, then the hour-glass field morphology might be caused by the magnetic field being forced from an initially cylindrically-symmetric morphology by the passage of the explosive outflow through the filament.  The BN/KL outflow is a very strong explosive outflow \citep{thaddeus1972}, the apparent origin of which coincides with the centre of the BN/KL clump.  The BN/KL outflow is one of the most energetic outflows known in a star-forming region, with a total energy in the outflow of $\sim 4\times 10^{40}$\,J \citep{kwan1976}. The outflow has a wide opening angle, and high-velocity wings with multiple ejecta, often referred to as the `bullets of Orion' \citep{allen1993}.  The sources BN, $n$ and $I$, located in the core of the BN/KL object, have proper motions consistent with their having undergone a close dynamical interaction $\sim 500\,$years ago \citep{gomez2005}.  It has been hypothesized that the BN/KL outflow was produced as a result of this interaction \citep{bally2005}.  This hypothesis is supported by the dynamic age of the BN/KL outflow, $\sim 500\,$years, which is comparable to the time since the interaction \citep{zapata2009}, and by the kinetic energy released by the interaction, $\sim 2\times 10^{40}\,$J \citep{gomez2005}, which is comparable to the energy in the outflow \citep{kwan1976}.

\begin{figure*}
\centering
\includegraphics[width=0.55\textwidth]{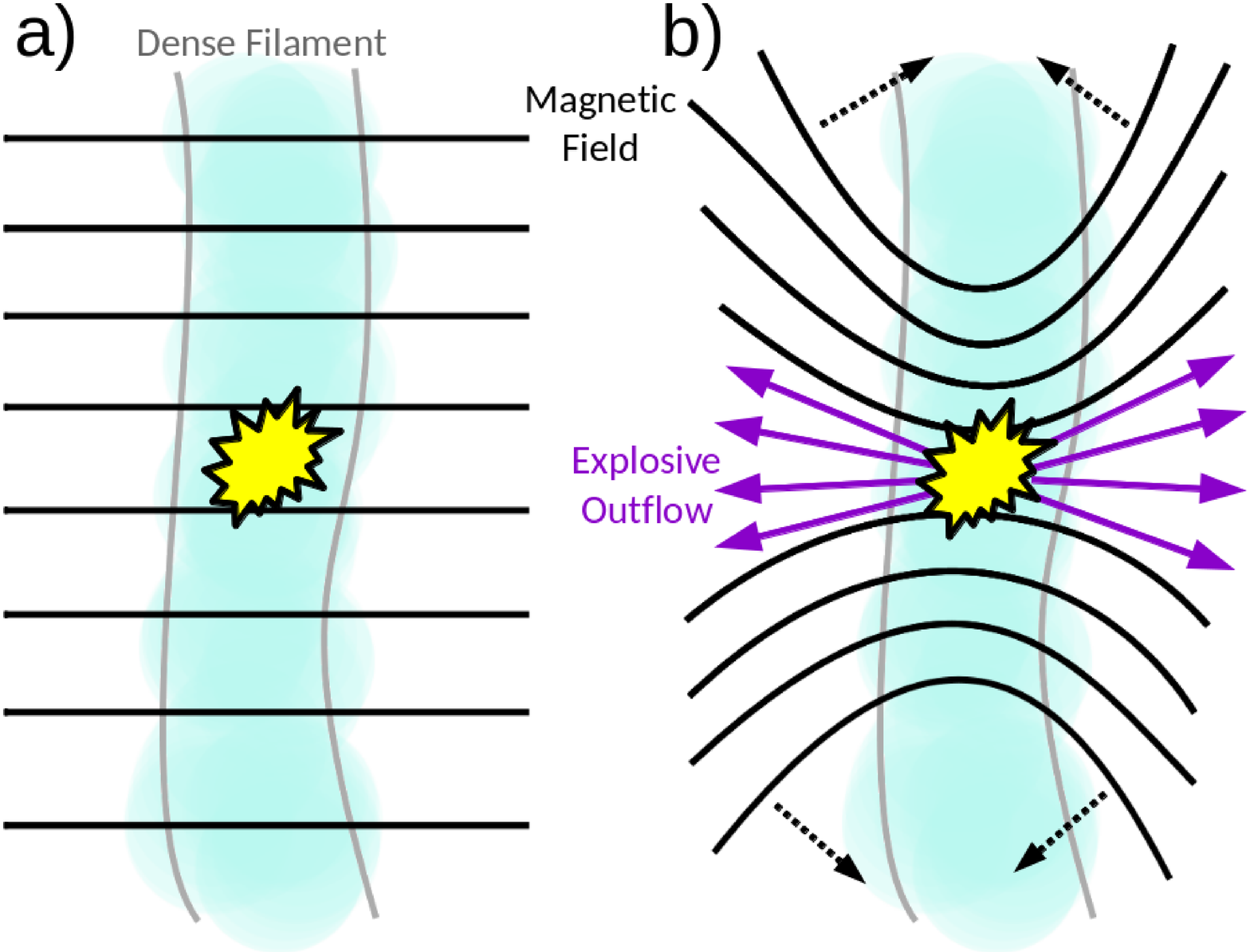}
\caption{A cartoon of another of our proposed formation mechanisms for the hour-glass magnetic field morphology of OMC~1, in which the field is shaped by the effects of the BN/KL outflow.  An initially cylindrically-symmetric field is disrupted by the explosive outflow.  The opening angle of the outflow determines, and approximately matches, the opening angle of the hour-glass of the magnetic field.}
\label{fig:cartoon_outflow}
\end{figure*}

Hence, one possible explanation for the field line orientation around the OMC~1 filament is that it started out in a cylindrically-symmetric configuration, perpendicular to the filament, and was subsequently distorted into its current configuration by the energetic outflow from the BN/KL object, the major axis and opening angle of which is approximately coincident with the orientation of the magnetic field hour-glass geometry.  {The orientation of the hourglass is ${-64.2\pm 6.5}$ degrees, measured east of north \citep{wardthompson2017}.  We estimate a position angle of the BN/KL outflow of $\sim - 61$ degrees from the visual extinction data presented by \citet{youngblood2016} (see their Figure 3), consistent with the orientation of the hourglass magnetic field.  The BN/KL outflow and the magnetic field morphology are compared in Figure~\ref{fig:outflow}.}  This formation mechanism is illustrated in Figure~\ref{fig:cartoon_outflow}.

We test these two hypotheses for the formation of the hour-glass morphology by considering the total energy and energy density in the Orion BN/KL region due to the magnetic field, the gravitational interaction of BN/KL and S, and the BN/KL outflow.

\subsection{Magnetic energy density of OMC~1}

The magnetic energy density is given by
\begin{equation}
U_{B}=\frac{B^{2}}{2\mu_{0}}
\label{eq:mag_energy_density}
\end{equation}
in SI units, where $\mu_{0}$ is the permeability of free space.  For {${B_{\rm pos}=\bfield}$}\,mG, {${U_{B}\sim\magU\times 10^{-7}}$\,J\,m${^{-3}}$}.  The total magnetic energy is then
\begin{equation}
E_{B}=U_{B}V,
\end{equation}
where $V$ is the volume over which the magnetic field is applied.  For our defined volume of OMC~1, {${E_{B}\sim\magE\times 10^{40}}$}\,J.

\subsection{Mass-to-Flux ratio in OMC~1}

We determine the mass-to-magnetic-flux ratio $M/\Phi$ in OMC~1 in units of the critical ratio,
\begin{equation}
\lambda = \frac{(M/\Phi)_{observed}}{(M/\Phi)_{critical}},
\end{equation}
where the observed $M/\Phi$ ratio is given by
\begin{equation}
\left(\dfrac{M}{\Phi}\right)_{obs}=\dfrac{\mu m_{\textsc{h}}N({\rm H}_{2})}{B},
\end{equation}
and the critical $M/\Phi$ ratio by
\begin{equation}
\left(\dfrac{M}{\Phi}\right)_{crit} = \dfrac{1}{2\pi\sqrt{G}}
\end{equation}
{\citep{nakano1978}. We note that the constant ${1/2\pi}$ is model-dependent and varies with source geometry (e.g. \citealt{mckee1993}), but should be correct to within a factor of a few.} The critical ratio is determined as described by \citet{crutcher2004}:
\begin{equation}
\lambda_{obs} = 7.6\times 10^{-21}\frac{N({\rm H}_{2})}{B_{\rm pos}},
\end{equation}
where $N({\rm H}_{2})$ is in units of cm$^{-2}$ and $B_{\rm pos}$ is in units of $\upmu{\rm G}$.  A value of $\lambda < 1$ (magnetically subcritical) indicates that the magnetic field strength is sufficiently high to support against gravitational collapse, while $\lambda>1$ (magnetically supercritical) indicates that the magnetic field cannot prevent gravitational collapse.  \citet{crutcher2004} further note that statistically, the observed $M/\Phi$ ratio will over-estimate the true value by a factor of 3, and so,
\begin{equation}
\lambda = \frac{1}{3}\lambda_{obs}.
\end{equation}
Note that this is a correction for the geometrical effect of overestimation of ${N(H_{2})}$ due to the unknown orientation of the source relative to the plane of the sky.

Using our best estimate of plane-of-sky magnetic field strength, {${B_{\rm pos}=\bfield}$}\,mG and our median column density value, $N({\rm H}_{2})=\coldens\times10^{23}$\,cm$^{-2}$, we find ${\lambda_{obs}\sim\mphi}$.  If the statistical correction given by \citet{crutcher2004} applies in OMC~1, this implies ${\lambda\sim\mphicorr}$.  However, as the OMC~1 filament appears to lie in or near the plane of the sky, applying this correction may cause us to significantly overestimate the degree to which OMC~1 is magnetically subcritical.  We thus use the observed value, ${\lambda_{obs}\sim\mphi}$, noting that this may be a slight overestimate.

A value of ${\lambda_{obs}\sim\mphi}$ suggests that the OMC~1 region is typically somewhat magnetically sub-critical, and so suggests that the magnetic field can provide support against gravitational collapse (i.e. the filament fragmenting or collapsing toward its axis) on the scales which we probe with these observations.  Although our spatially-averaged value of $\lambda_{obs}$ is less than unity, it is clear that OMC~1 cannot be magnetically sub-critical everywhere, as the region is an active site of star formation, and so at least some parts of the cloud must have undergone gravitational collapse in the past.  This result suggests that on the scales probed by our observations, OMC~1 is at or near magnetic criticality.  We discuss the gravitational stability of OMC~1 further in Section~\ref{sec:grav_stab}.

These values are comparable to the ratios of magnetic to gravitational force measured in OMC~1 by \citet{koch2014}, who found a ratio of $\lambda\sim 0.75$ using Caltech Submillimeter Array (CSO) observations, and of $\lambda\sim 1.26$ using Submillimeter Array (SMA) observations.  Both of these values are consistent with unity, again suggesting that OMC~1 is near magnetic criticality.

\subsection{Gravitational potential energy of OMC~1}

We determine the masses of the BN/KL and S clumps using the column density map described in Section~\ref{sec:vol_density}.  Measuring the extent and central positions of the BN/KL and S clumps from our column density map, we estimate a mass of $1001\pm791$\,M\sun\ for the BN/KL clump and a mass of $286\pm226$\,M\sun\ for the S clump, with a plane-of-sky separation of 88\,arcsec, equivalent to 0.166\,pc at a distance of 388\,pc.  We assume that all of the mass along the line of sight towards each clump is associated with that clump, and hence that any mass distributed along the line of sight is negligible.  Both clumps are significantly extended objects, with the BN/KL clump having major and minor axis diameters of $\sim$1.5\,arcmin and $\sim$1.0\, arcmin respectively, while the S clump has major and minor axis diameters of $\sim$1.0\,arcmin and 0.7\,arcmin respectively.  We determine a total mass in the area of OMC~1 over which we performed our CF analysis of ${1413\pm1116}$\,M\sun.  The large majority (${\sim 92}$\%) of the mass in the center of OMC~1 is thus in the BN/KL and S clumps, suggesting that our assumption that BN/KL and S dominate the mass distribution along their lines of sight is justified.

\subsubsection{Gravitational stability of the OMC~1 filament}
\label{sec:grav_stab}

{We first estimated the global gravitational stability of the OMC~1 region using the \citet{ostriker1964} critical mass per unit length (line mass) for an isothermal filament,}
\begin{equation}
{\left(\frac{M}{L}\right)_{crit} = \frac{2\sigma}{G},}
\end{equation}
{where ${\sigma}$ represents the gas velocity dispersion.  Assuming initially that the gas in OMC~1 is supported by thermal pressure, for a typical gas temperature of 15\,K (representative of the conditions we measure in OMC~1), the critical line mass is ${(M/L)_{crit,15\,{\rm K}}=1.62\times 10^{15}}$\,kg\,m${^{-1}\approx 25}$\,M\sun\,pc${^{-1}}$.  As discussed above, we measured a total mass of ${1413\pm1116}$\,M\sun\ over a 0.35\,pc length of the OMC~1 region.  Thus, in the vicinity of OMC~1, we measure a line mass of ${(M/L)=4038\pm3190}$\,M\sun\,pc${^{-1}}$, significantly larger than the thermal critical line mass.  This suggests that the OMC~1 filament, in this region, would be significantly gravitationally unstable in the absence of either turbulent support or a magnetic field.}

{If we assume that the non-thermal gas velocity dispersion acts as a hydrostatic pressure in providing support against gravitational collapse (the microturbulent assumption; c.f Chandrasekhar 1951a,b\nocite{chandrasekhar1951a}\nocite{chandrasekhar1951b}), we can take ${\sigma=\sigma_{v,\textsc{3D}}=\sigma_{v}\sqrt{3}=2.30}$\,km\,s${^{-1}}$.  Note that this assumes that the velocity dispersion is isotropic.  We then find a critical line mass ${(M/L)_{crit,turb}=1.59\times 10^{17}}$\,kg\,m${^{-1}\approx 2470}$\,M\sun\,pc${^{-1}}$, comparable to but slightly lower than our observed line mass.  This is likely to represent an upper limit on the amount of support which can be provided by turbulent gas pressure.  These results suggest that the OMC~1 filament can at best be marginally supported against collapse by turbulent gas pressure.  We also note that previous studies of the integral filament have found its radial density profile to be inconsistent with the \citet{ostriker1964} self-gravitating isothermal cylinder model (\citealt{johnstone1999}; \citealt{salji2015a}).}

{\citet{fiege2000} modified the \citet{ostriker1964} stability criterion to estimate the stability of magnetized filaments, proposing the criterion}
\begin{equation}
{\left(\frac{M}{L}\right)_{crit,mag} = \left(\dfrac{M}{L}\right)_{crit}\left(1-\dfrac{\mathcal{M}}{\lvert\mathcal{W}\rvert}\right)^{-1},}
\end{equation}
{where ${\mathcal{M}}$ is the magnetic energy per unit length, and ${\mathcal{W}}$ is the gravitational energy per unit length,}
\begin{equation}
{\mathcal{W} = -\left(\frac{M}{L}\right)^{2}G.}
\end{equation}

Using our total mass of 1413\,M\sun, we estimate a gravitational potential energy per unit length of ${\lvert\mathcal{W}\rvert = 4.52\times 10^{24}}$\,J\,m${^{-1}}$.  This is equivalent to a total gravitational potential energy in OMC~1 of ${E_{G}=4.9\times 10^{40}}$\,J, and so to a gravitational potential energy density ${U_{G}=1.9\times 10^{-7}}$\,J\,m${^{-3}}$ over the volume over which we performed our CF analysis (${V=\pi r^{2}L=2.62\times10^{47}}$\,m${^{-3}}$), very similar to our estimated magnetic energy density for the region.

We estimate the magnetic energy per unit length to be ${\mathcal{M}=4.44\times 10^{24}}$\,J\,m${^{-1}}$.  The magnetic critical line mass is then given, in the 15\,K case, by $(M/L)_{crit,mag,15\,{\rm K}}=9.56\times 10^{16}$\,kg\,m$^{-1}=1480$\,M\sun\,pc$^{-1}$.  In the turbulent-support case, the magnetic critical line mass is  $(M/L)_{crit,mag,turb}=9.35\times 10^{18}$\,kg\,m$^{-1}=1.45\times10^{5}$\,M\sun\,pc$^{-1}$. 

{These values suggest that the magnetic field contributes significantly to supporting the filament against gravitational collapse.  In the thermal case, our results suggest that the filament is marginally gravitationally unstable, although the critical and observed values of ${M/L}$ match within experimental uncertainty.  In the turbulent case, we find that the filament is definitively stable, and supported by its magnetic field.  The former of these scenarios -- a filament in approximate equilibrium between gravitational collapse and magnetic support -- is more physically plausible than the latter, particularly as the significant deviation from cylindrical symmetry in the magnetic field suggests that the field has been significantly deviated in the recent past.  If the filament has collapsed gravitationally, thereby compressing the local magnetic field and so evolving to a state of approximate equilibrium, we would expect an observed line mass similar to, rather than significantly smaller than, the critical line mass.}

{Our results suggest that the turbulence in OMC~1 is not providing significant support against gravitational collapse.  This is not a surprising result; the microturbulent assumption holds only on scales smaller than the thermal Jeans length (see \citealt{maclow2004}, and references therein).  The thermal Jeans length (${\lambda_{J}=c_{s}\sqrt{\pi/G\rho}}$; \citealt{jeans1928}) in OMC~1 is ${\lambda_{J}\sim 0.03}$\,pc for our representative values of ${T=15}$\,K and ${n=\voldens\times10^{6}}$\,cm${^{-3}}$.  Thus, while turbulence may provide some support against gravitational collapse on small scales, large-scale gravitational motions in OMC~1 (occurring on size scales ${\gtrsim 10^{-1}}$\,pc) cannot be supported against in this manner.}

{Our results therefore suggest that the gravitational and magnetic energy densities in OMC~1 are similar.  However, the analysis above is performed for a uniform cylindrically-symmetric geometry, which is demonstrably not the case in OMC~1.  As the large majority of the mass of OMC~1 is within the BN/KL and S clumps, we estimate the gravitational potential energy density of the BN/KL-S system as a check on our results.  We thus proceed by assuming that the gravitational potential of the region is currently dominated by these two clumps, regardless of their formation mechanism.}  We determine the gravitational potential energy of the BN/KL-S system in two limits: firstly, by considering the clumps as separate point sources, and secondly, by considering the system as a uniform-density prolate spheroid. 

\subsubsection{Point-source model}

As we do not know the line-of-sight component of the separation between the two clumps, we multiply our measured separation of 0.166\,pc by $\sqrt{2}$ (assuming {conservatively} that the filament is orientated at 45 degrees to the plane of the sky), and so estimate a total separation between the two clumps of $\sim 0.23$\,pc.  From these values we infer a gravitational potential energy in OMC~1 using the relation
\begin{equation}
E_{G}=-\frac{GM_{1}M_{2}}{r},
\label{eq:gpe_point}
\end{equation}
were $M_{1}$ is the mass of the BN/KL clump, $M_{2}$ is the mass of the S clump and $r$ is the separation of the clumps.

Using equation~\ref{eq:gpe_point}, we find $E_{G}=-1.0\times 10^{40}$\,J.  This is comparable to our estimate of magnetic energy in OMC~1 {and to our estimate of $E_{G}$ in Section~\ref{sec:grav_stab}}.  However, our estimate of the total magnetic energy of OMC~1 is determined by multiplying the mean magnetic energy density by a larger volume of OMC~1 than is occupied by the BN/KL-S system.  In order to make a more meaningful comparison, we compare the magnetic energy density in OMC~1 to the gravitational potential energy density in a region just enclosing the BN/KL-S system: a box of angular width 1$^{\prime}$15$^{\prime\prime}$ and angular height 2$^{\prime}$50$^{\prime\prime}$, equivalent to 0.141\,pc and 0.320\,pc respectively at a distance of 388\,pc.  On the assumption that the OMC~1 filament is cylindrical and inclined at 45 degrees to the plane of the sky, we infer a volume occupied by the BN/KL-S system of $2.1\times 10^{47}$\,m$^{3}$, and so a gravitational potential energy density,
\begin{equation}
U_{G}=\frac{|E_{G}|}{V},
\end{equation}
of $U_{G}=0.5\times10^{-7}$\,J\,m$^{-3}$, a value comparable to our representative magnetic energy density, $U_{B}\approx1.0\times10^{-7}$\,J\,m$^{-3}$.

\subsubsection{Prolate-spheroid model}

We also model the BN/KL-S system as a uniform-density prolate spheroid with total mass 1286\,M\sun\ (the combined masses of BN/KL and S), semimajor axis 0.16\,pc and semiminor axes 0.071\,pc.  We calculate the gravitational potential energy using the relation
\begin{equation}
E_{G}=-\frac{8}{15}\pi^{2}G\rho^{2}a_{1}^{4}a_{3}\times\frac{1}{e}\ln\left(\frac{1+e}{1-e}\right),
\label{eq:gpe_spheroid}
\end{equation}
where $a_{1}$ is the semiminor axis, $a_{3}$ is the semimajor axis, $\rho$ is the density of the spheroid, defined as the total mass divided by $\frac{4}{3}\pi a_{1}^{2}a_{3}$, and $e$ is the eccentricity of the spheroid,
\begin{equation}
e=\sqrt{1-\left(\frac{a_{1}}{a_{3}}\right)^{2}}.
\end{equation}
See, e.g., \citet{binney2008} for a derivation of this result.

Using equation~\ref{eq:gpe_spheroid}, we find $E_{G}=-8.6\times 10^{40}$\,J.  This is comparable to the total magnetic energy which we estimate for OMC~1, {and to our previous estimates of $E_{G}$}.  Dividing this value by the volume of the spheroid as defined above, we find a gravitational potential energy density of $U_{G}=8.8\times10^{-7}$\,J\,m$^{-3}$, a value somewhat larger than, but comparable to, our estimated magnetic energy density.

\subsection{Energy density of the BN/KL outflow}

The total energy in the BN/KL outflow is $\sim 4\times 10^{40}$\,J \citep{kwan1976}.  We estimate a mean energy density in the outflow by assuming that both wings of the outflow occupy equal volumes, each a sector of a sphere with an opening angle of 1 radian (estimated from the data presented by \citealt{bally2015}) and a radius of 0.26\,pc, the furthest distance in projection travelled by a Herbig-Haro object associated with the outflow (\citealt{bally2015}; correcting for their assumption of a distance of 414\,pc to OMC~1).  The total volume of the outflow is then:
\begin{equation}
V_{\rm outflow}=2\times\frac{2\pi r^{3}}{3}\left[1-\cos(\phi)\right],
\label{eq:outflow_vol}
\end{equation}
where $\phi$ is the half-angle of the outflow.  For the values given above, $V_{\rm outflow}=2.7\times10^{47}$m$^{3}$.  The mean energy density of the BN/KL outflow would then be $U_{\rm outflow}\sim 1.5\times10^{-7}$\,J\,m$^{-3}$, comparable to the energy density which we infer for the magnetic field in OMC~1.  However, it must be noted that the energy of the BN/KL outflow will not be evenly distributed within the volume defined by equation~\ref{eq:outflow_vol}: the `bullets of Orion', which occupy the majority of the volume under consideration, are Herbig-Haro objects ejected ballistically by the outflow, and have a current total kinetic energy of $\sim 10^{37}$\,J \citep{allen1993}.  The large majority of the energy of the outflow is concentrated in the central, highly-collimated outflow that caused the ejection of the `bullets'.  If we assume that the volume occupied by the collimated outflow is negligible compared to the volume occupied by the bullets, then we find an energy density for the ballistically-ejected bullets of $U_{\rm bullets}\sim 4\times10^{-11}$\,J\,m$^{-3}$.  We use the former value in the subsequent discussion, as representing an upper limit on the energy density of the large-scale outflow.

\subsection{Alfv\'{e}n velocity in OMC~1}

We calculated the Alfv\'{e}n velocity $c_{A}$ in OMC~1 using the relation
\begin{equation}
c_{A}=\frac{B}{\sqrt{\mu_{0}\rho}},
\end{equation}
where all symbols are as defined above.  For our representative density of {${\voldens\pm\voldenserr\times 10^{6}}$}\,cm$^{-3}$ and field strength of {${B_{\rm pos}=\bfield\pm\bfielderr}$}\,mG, we infer an Alfv\'{e}n velocity of {${\alfven\pm\alfvenerr}$}\,km\,s$^{-1}$.

From this value we can calculate the maximum distance that the magnetic field could have deviated Alfv\'{e}nically from its original configuration in 500 years (the approximate age of the BN/KL outflow; \citealt{gomez2005}), and find that the maximum deviation is {${\alfvendev\pm\alfvendeverr\times10^{-3}}$}\,pc.  This value is orders of magnitude smaller than the size scale on which we see variation in the geometry of the magnetic field ($\sim 10^{-1}$\,pc).  We discuss this result further in Section~\ref{sec:balance}.

\subsection{Kinetic energy in OMC~1}

\subsubsection{Kinetic energy of the BN/KL-S interaction}
\label{sec:los_ke}

We calculate the kinetic energy of the relative line-of-sight motion of Orion BN/KL and S, in order to determine whether the energy of the clumps' relative motion could significantly affect the energy balance of the region.  We determine line-of-sight velocities from our fitting of the HARP C$^{18}$O data \citep{buckle2010}.  We measure average systemic velocities of $v_{{\rm los,BN}}= 8.8\pm0.8$\,km\,s$^{-1}$ for BN/KL and $v_{{\rm los,S}}= 6.8\pm0.3$\,km\,s$^{-1}$ for S, and hence a relative velocity between the clumps of $v_{\rm rel,los}=2.0\pm0.9$\,km\,s$^{-1}$.

Assuming that in the inertial frame of the BN/KL-S system the clumps began their motion from rest, we can deduce from the conservation of linear momentum that
\begin{align}
  v_{f,{\rm BN}}=\frac{M_{{\rm S}}}{M_{{\rm BN}}+M_{{\rm S}}} v_{\rm rel,los} \label{eq:lin_mom_1} \\ 
  v_{f,{\rm S}}=-\frac{M_{{\rm BN}}}{M_{{\rm BN}}+M_{{\rm S}}} v_{\rm rel,los}, \label{eq:lin_mom_2}
\end{align}
where $v_{f}$ is the line-of-sight velocity of the clump in the inertial frame of the BN/KL-S system, and $M_{{\rm BN}}$ and $M_{{\rm S}}$ are the masses of BN/KL and S as determined above.  From equations~\ref{eq:lin_mom_1} and \ref{eq:lin_mom_2} we determine line-of-sight velocities of $v_{f,{\rm BN}}\sim 0.4$\,km\,s$^{-1}$ and $v_{f,{\rm S}}\sim -1.6$\,km\,s$^{-1}$.  Using our previous mass estimates for BN/KL and S and the equation for translational kinetic energy,
\begin{equation}
  E_{K,trans}=\frac{1}{2}Mv^{2},
\end{equation}
we find a total line-of-sight kinetic energy of $\sim 1.1\times10^{38}\,$J for BN/KL and $\sim 3.9\times10^{38}\,$J for S, two orders of magnitude lower than the gravitational, magnetic and outflow energies.  It should be noted that this is the energy of only one of the three components of the relative motion of BN/KL and S.  However, the kinetic energy of the motion of the clumps in the plane-of-sky directions would have to be $\sim 10^{2}$ times that of the motion along the line of sight -- i.e. the plane-of-sky velocities would have to be $\sim 10$ times the line-of-sight velocities -- to significantly affect the energy balance of the region.

\subsubsection{Internal thermal energy of BN/KL and S}
\label{sec:int_ke}

We calculate the internal thermal energies of Orion BN/KL and S by determining average temperatures for each core using the temperature map described in Section~\ref{sec:vol_density}.  We measure a mean temperature of $15\pm3\,$K in BN/KL, and of $17\pm3\,$K in S.  The internal thermal energy is given by
\begin{equation}
  E_{K,thermal}=\frac{3}{2}Mc_{s}^{2},
\end{equation}
where $c_{s}$ is the sound speed in the gas,
\begin{equation}
  c_{s}=\sqrt{\frac{k_{\textsc{b}}T}{m}}.
\end{equation}
For a typical core temperature of 15\,K and the masses of BN/KL and S as determined above, we find a sound speed $c_{s}=0.23$km\,s$^{-1}$ and a total thermal kinetic energy for BN/KL and S of $2.0\times 10^{38}$\,J, insufficient to significantly affect the energy balance of the region.

\subsubsection{Internal non-thermal energy of BN/KL and S}


The internal non-thermal kinetic energy is given by
\begin{equation}
E_{K,non-thermal}=\frac{3}{2}M\sigma_{v,\textsc{nt}}^{2}.
\end{equation}
For the internal non-thermal linewidth {${\sigma_{v}=1.32}$\,km\,s${^{-1}}$} and the masses of BN/KL and S as determined above, the total non-thermal kinetic energy of BN/KL and S is $\sim 6.6\times10^{39}$\,J.  This is slightly lower than, but comparable to, the lower end of our estimated range of gravitational energies.  This {would suggest} that the non-thermal kinetic energy may contribute to the total energy balance of OMC~1, but does not dominate it.  {However, as discussed in Section~\ref{sec:grav_stab} above, it is likely that non-thermal motions in OMC~1 are not providing significant support against gravitational collapse on scales larger than the Jeans length, ${\lambda_J \sim 0.03}$\,pc.}

\section{Discussion}
\label{sec:balance}

\begin{table}
\centering
\caption{Properties of OMC~1 relevant to the energy balance of the region.}
\begin{tabular}{cc}
\toprule
Property & Value \\
\midrule
$U_{B}$ & {${\sim\magU\times10^{-7}}$}\,J\,m$^{-3}$\\
$U_{G}$ & $0.5$-$8.8\times10^{-7}$\,J\,m$^{-3}$ \\
$U_{outflow}$ & $\sim 1.5\times10^{-7}$\,J\,m$^{-3}$ \\
$U_{bullets}$ & $\sim 4\times10^{-11}$\,J\,m$^{-3}$ \\
\rule{0pt}{4ex}
$E_{B}$ & {${\sim\magE\times 10^{40}}$}\,J \\
$E_{G}$ & $-(1.0$-$8.6)\times 10^{40}$\,J \\
$E_{outflow}$\tablenotemark{a} & $\sim 4\times10^{40}$\,J \\
$E_{bullets}$\tablenotemark{b} & $\sim 10^{37}$\,J \\
$E_{K,trans}$ & $\sim 5.0 \times 10^{38}$\,J \\
$E_{K,thermal}$ & $\sim 2.0\times 10^{38}$\,J \\
$E_{K,non-thermal}$ & $\sim 6.6\times10^{39}$\,J \\
\bottomrule
\end{tabular}\\
$^{\rm a}$\citet{kwan1976}\\
$^{\rm b}$\citet{allen1993}
\label{tab:energetics}
\end{table}

\subsection{Interaction of the magnetic field and the BN/KL outflow}

The magnetic field in OMC~1 is clearly highly ordered, despite the presence of the highly energetic BN/KL outflow, which suggests that, on large scales, the magnetic field is sufficiently strong not to be totally disrupted by the outflow.  Our estimates of the magnetic and outflow energy densities are very similar, although both are probably correct only to within an order of magnitude.  The key quantities relevant to the energetics of OMC~1 are summarized in Table~\ref{tab:energetics}.

The central, highly-collimated, part of the BN/KL outflow is likely to have sufficient energy to disrupt the local magnetic field.  \citet{tang2010} showed, using 870-$\upmu$m SMA observations with 1-arcsec resolution, that the polarization half-vectors at the centre of the BN/KL region trace an approximately circular structure, and proposed that this might be due to the local magnetic field being dragged along by the outflow.  The effect of this circular polarization structure on our maps is to produce a completely depolarized region approximately the size of the JCMT beam (FWHM $\sim 14$\,arcsec) at the central position of the BN/KL outflow, consistent with observations by \citet{schleuning1998}, \citet{rao1998} and \citet{houde2004}.

As discussed above, the Alfv\'{e}n velocity in OMC~1 is sufficiently small that the distortion in the magnetic field, which extends significantly beyond the maximum extent of the outflow (on size scales $\sim 10^{-1}\,$pc), cannot have occurred through an Alfv\'{e}nic perturbation of the field in the 500 years that the BN/KL outflow has existed.  A perturbation in the magnetic field expanding Alfv\'{e}nically could have deviated the magnetic field on a maximum scale $\sim 10^{-3}$\,pc in 500 years.  However, outflows and ejecta moving supersonically and super-Alfv\'{e}nically could alter the magnetic field more rapidly, through compression or dragging of gas into which the magnetic field is frozen (e.g. \citealt{padoan1999}).  The maximum deviation of the field would thus be set by the maximum travel distance of the outflow ejecta.

Estimates of the typical line-of-sight velocity of the outflow ejecta range from $\sim 80$\,km\,s$^{-1}$ (e.g. \citealt{furuya2009}) to $\sim 150$\,km\,s$^{-1}$ (e.g \citealt{bally2017}), significantly greater than the Alfv\'{e}n velocity.  Ejecta travelling at a constant velocity of $150$\,km\,s$^{-1}$ could travel a maximum distance of 0.077\,pc.  Although some deceleration of the ejecta over time is likely, the maximum travel distance of the ejecta is $\sim 10^{-2}$\,pc, an order of magnitude smaller than the size scale of the deviations in the magnetic field.  Inspection of Figures~\ref{fig:outflow} and \ref{fig:oriona} shows that the maximum extent of the deviation in the magnetic field is significantly larger than the maximum extent of the outflow.

Moreover, while the total energy densities of the magnetic field and of the BN/KL outflow are comparable, the energy density of the ballistic outflow ejecta is several orders of magnitude smaller than that of the magnetic field.  Thus we conclude that while there is sufficient energy in the BN/KL outflow to potentially alter the geometry of the magnetic field in OMC~1, the outflow is too young to have caused the large-scale hour-glass shape seen in the magnetic field in OMC~1.

It hence seems plausible that the direction of propagation, and the opening angle, of the ballistically ejected BN/KL outflow (the `bullets'), may be constrained by the magnetic field morphology in the region; i.e. on large scales the outflow is being shaped by the magnetic field, rather than the converse.

\subsection{Interaction of the magnetic field and the gravitational potential}

We now consider whether the hour-glass morphology could have been caused by {gravitationally-driven motion of material in the filament}.  The gravitational potential energy density and magnetic energy density of the central part of OMC~1 are comparable to one another, suggesting that the filament may be in or near equipartition of energy between the gravitational and magnetic fields, and may hence have been in approximate equilibrium before the formation of the BN/KL outflow.

If the primordial magnetic field were uniform and perpendicular to the filament, then its energy density ought to have been lower than that which we now observe; distortion of the magnetic field by {gravitationally-driven motions, either of material along to filament to form the BN/KL and S clumps, or of the BN/KL and S clumps themselves,} might have compressed the field lines and so increased the magnetic field strength and the magnetic energy density.

We {therefore} hypothesize that {the the magnetic field has been compressed by the large-scale motions of material along the filament} to the point that its energy is now comparable to that of the gravitational interaction of the two clumps in OMC~1, and hence that any motion of the clumps toward one another {has} been halted or slowed by the balance of forces between the gravitational interaction and the magnetic field, with the magnetic field providing {a `cushion'} preventing further {flow of gas along the filament} (see Figure~\ref{fig:cartoon_gravity}).  {Any further interaction of the BN/KL and S clumps will thus be secular and mediated by ambipolar diffusion.}

\subsection{Comparison with existing measurements}

The magnetic field strength which we infer in OMC~1 is very strong, but not unprecedentedly so.  The mG-strength field which we observe in OMC~1 has large-scale structure that varies on size scales $\sim 10^{-1}$\,pc, which we observe at a spatial resolution of $0.026$\,pc (for our assumed distance of 388\,pc).  Magnetic field strengths of the order of a few mG have been measured in dense gas in high-mass star-forming regions on a wide variety of spatial scales.  For example, \citet{curran2007}, observing with SCUPOL at 14-arcsec resolution, measured a magnetic field strength of 5.7\,mG in Cepheus A (0.05\,pc spatial resolution for their assumed distance of 725\,pc), and magnetic field strengths $\sim 1$\,mG in both DR21(OH) (2\,pc spatial resolution at 3\,kpc) and the low-mass star-forming region RCrA (0.009\,pc spatial resolution at 130\,pc).  Recent ALMA observations have found magnetic field strengths in the range 0.2--9\,mG in the high-mass W43-MM1 star-forming region, with 0.5-arcsec (${\sim 0.01}$\,pc) resolution \citep{cortes2016}, and field strengths of 0.4--1.7\,mG have recently been measured in DR21 using SMA observations with $3-4$-arcsec resolution (${0.02-0.03}$\,pc for their assumed distance of 1.4\,kpc) \citep{ching2017}.  Other measurements of mG-strength magnetic fields include (but are not limited to): Girart et al. (2009, 2013)\nocite{girart2009}\nocite{girart2013}; \citet{crutcher2010}, and references therein; \citet{stephens2013}; Qiu et al. (2013, 2014)\nocite{qiu2013}\nocite{qiu2014}; Pillai et al. (2015, 2016)\nocite{pillai2015}\nocite{pillai2016}.

There are various existing measurements of the magnetic field strength in the OMC~1 molecular cloud.  \citet{hildebrand2009} used Hertz data with 20\,arcsec resolution to estimate a plane-of-sky magnetic field strength in the OMC~1 region of 3.8\,mG (without formal uncertainties), using their `dispersion function' method of measuring the dispersion in angle of the magnetic field due to turbulence.  The field measured using the `unsharp-masking' method presented in this work is approximately consistent with the field strength estimated by \citet{hildebrand2009}.

\citet{crutcher1999} measured the CN Zeeman effect at two positions in the northern bright peak of OMC~1, R.A. (J2000) $= 05^{h}35^{m}14^{s}.5$ Dec. (J2000) $= -05^{\circ}22^{\prime}06^{\prime\prime}.5$ and R.A. (J2000) $= 05^{h}35^{m}13^{s}.5$ Dec. (J2000) $= -05^{\circ}22^{\prime}51^{\prime\prime}.5$, detecting a line-of-sight magnetic field strength of $-0.36\pm0.08$\,mG at the northern position and making no detection at the southern position.  The CN Zeeman effect is thought to measure the line-of-sight magnetic field strength in molecular clouds at densities $10^{5}-10^{6}$\,cm$^{-3}$ \citep{crutcher1996}, comparable to the densities which we consider in this work.  This would suggest that the line-of-sight magnetic field strength is an order of magnitude lower than the plane-of-sky field strength.  However, if the hour-glass morphology of the magnetic field in OMC~1 is three-dimensional, and rotationally symmetric about the main axis of the OMC~1 filament, and the OMC~1 filament is orientated in or near the plane of the sky, then the sum of the line-of-sight components of the magnetic field strength vectors at any given position ought to cancel, and so the measured line-of-sight magnetic field strength ought to be significantly smaller than the plane-of-sky field strength.  If this is the case then our results are not necessarily inconsistent with those of \citet{crutcher1999}.

\citet{houde2009} measured a magnetic field strength in OMC~1 of 0.76\,mG using SHARP data at 12\,arcsec resolution, arguing that the integration of polarized emission along the line of sight of the molecular cloud and within the beam of the telescope leads to overestimation of the magnetic field strength.  They account for this effect by attempting to infer the turbulent correlation length of the cloud.  We discuss this effect further below.

Interferometric observations of OH maser emission in the BN/KL region consistently produce milli-Gauss magnetic field strengths.  \citet{cohen2006} measured magnetic field strengths in the range 1.8 to 16.3 mG from MERLIN observations at ${\sim 0.15}$\,arcsec resolution, and suggested a general line-of-sight magnetic field strength of ${\sim 1-3}$\,mG, within which there are localised regions of higher field strength.  \citet{hansen1983} found a magnetic field strength ${\sim 3}$\,mG from 0.2-arcsec VLA observations.  \citet{norris1984}, observing with MERLIN at 0.3-arcsec resolution, also found a magnetic field strength ${\sim 3}$\,mG, while \citet{johnston1989} estimated a magnetic field strength of ${1-3}$\,mG based on 0.3-arcsec resolution VLA observations.

\citet{tang2010} determine a magnetic field strength $\geq 3$\,mG in Orion BN/KL.  They argue that dense clumps which they observe in NH$_{3}$ in the BN/KL region (with angular sizes $< 1$\,arcsec) are magnetically confined, and that this magnetic confinement requires a field strength $\geq 3$\,mG if it is to be maintained in the presence of the energetic outflows in this region.  This value is consistent with our measured magnetic field strength.

\subsection{Choice of $Q^{\prime}$ parameter}
\label{sec:q}

The choice of the normalisation parameter, {${Q^{\prime}}$}, in the CF equation (equation~\ref{eq:cf}) has been the subject of considerable debate in the literature.  The accuracy of the CF equation is affected by the integration of polarized emission both along the line of sight of the molecular cloud and within the beam of the telescope (e.g. \citealt{houde2009}).  Both of these averaging effects will, if the field is uncorrelated within the beam or between turbulent cells along the line of sight, cause the dispersion in angle to be underestimated, and so cause the magnetic field strength to be overestimated.

\citet{ostriker2001} determined that for realistic molecular cloud geometries, and where angular dispersion $\langle\sigma_{\theta}\rangle<25$ degrees, a normalisation parameter $Q^{\prime} \approx 0.5$ is required to accurately recover the plane-of-sky magnetic field strength.  This is an effect of cloud geometry, and is independent of smoothing effects.

\citet{heitsch2001} investigated the effect of smoothing their simulations of magnetized clouds (equivalent to observing with poorer resolution), and found that for strong magnetic fields with well-resolved angular field structure (as we have in Orion, with the hour-glass morphology), the CF method produces accurate results, typically correct to within a factor of 2.  \citet{heitsch2001} found that for very poorly-resolved and/or weak fields, the CF method could overestimate the magnetic field strength by up to a factor $\sim 10$, but none of these cases apply to OMC~1.  \citet{crutcher2004}, working with JCMT SCUPOL data (the same resolution as our own data and observing clouds at comparable distances), discussed these effects and suggested that for well-resolved filaments and cores, the \citet{ostriker2001} value of $Q^{\prime}\sim 0.5$ is appropriate, noting that it is accurate to $\sim 30\%$.

Modelling suggests that the number of independent turbulent eddies $N$ along the LOS causes the standard CF method to overestimate the magnetic field strength $B_{\rm pos}$ by a factor of $\sqrt{N}$ (e.g. \citealt{houde2009}; \citealt{cho2016}).  \citet{cho2016} proposed that the number of independent turbulent eddies along the line of sight can be estimated from the standard deviation of centroid velocities normalized by the average line-of-sight velocity dispersion over the region under consideration, i.e.
\begin{equation}
\frac{\sigma_{V_c}}{\sigma_{v}}\sim \frac{1}{\sqrt{N}},
\end{equation}
where $\sigma_{V_c}$ is the standard deviation of the mean of the centroid velocities measured across OMC~1, and $\sigma_{v}$ is the average of the line-of-sight velocity dispersions measured across OMC~1.  Using our C$^{18}$O data, we found that $\sigma_{v} = 1.33\pm 0.31$\,km\,s$^{-1}$ (see Section~\ref{sec:vel_disp}).  From the same data, measuring over the same area, we find a value of $\sigma_{V_c} = 0.97\pm 0.03$\,km\,s$^{-1}$.  Thus, we estimate a value of $1/\sqrt{N}=0.73\pm0.31$, slightly less than, but consistent with, unity.  This would suggest that there are few ($\lesssim2$) turbulent eddies along the line of sight.  Hence, any overestimation of $B_{\rm pos}$ resulting from LOS effects in our results should be small, and will not alter the order of magnitude of our measured magnetic field strength, or any of our scientific conclusions.

\section{Conclusions}

In this paper we have determined the magnetic field strength in the OMC~1 region using a Chandrasekhar-Fermi analysis of polarization observations made using the POL-2 polarimeter on the JCMT as part of the BISTRO survey and of POL-2 commissioning work.  We used archival SCUBA-2 and HARP observations in order to determine the volume density and gas velocity dispersion in OMC~1.  We estimated the angular dispersion in OMC~1 by applying a smoothing kernel to the distribution of angles and subtracting the smoothed magnetic field direction from the measured distribution of angles, a method analogous to unsharp masking.

We measured {${B_{\rm pos}/\sqrt{n}=\bfrootn\pm\bfrootnerr\,\upmu{\rm G}\,{\rm cm}^{-\frac{3}{2}}}$} in OMC~1, and hence for a typical gas density of {${n({\rm H}_{2})=\voldens\pm\voldenserr\times10^{6}}\,$}cm$^{-3}$, we determined a plane-of-sky magnetic field strength of {${B_{\rm pos}=\bfield\pm\bfielderr}$}\,mG, {where ${\delta B_{\rm pos}=\bfielderr}$\,mG represents a predominantly systematic uncertainty}.  This value is comparable to the magnetic field strength of 3.8 mG measured in OMC~1 by \citet{hildebrand2009}, {and to previous Zeeman measurements of OH masers in the BN/KL region}, and is comparable to magnetic field strengths measured in other high-mass star-forming regions.

The magnetic field in OMC~1 shows a distinctive hour-glass morphology.  We investigated the relative importance of the gravitational {instability of the filament and gravitational potential} of the Orion BN/KL and S clumps, and of the highly-energetic BN/KL outflow, in shaping the magnetic field in OMC~1.  We investigated the relative contribution of the magnetic field, gravitational interaction, and outflow to the energy balance in OMC~1.  We found that the magnetic field has an energy density {${\sim\magU\times 10^{-7}}$}\,J\,m$^{-3}$.  We estimated the gravitational potential energy density in the centre of OMC~1 to be of the order $10^{-7}$\,J\,m$^{-3}$, and the outflow energy density to be also $\sim 10^{-7}$\,J\,m$^{-3}$ (although we expect the energy density to be significantly non-uniform across the volume of the outflow).  Hence, we expect each of these effects to contribute similarly to the energy balance in OMC~1.

We investigated the translational, thermal and non-thermal kinetic energies in OMC~1, and found them to be smaller than the other terms contributing to the energy balance of OMC~1.  The non-thermal kinetic energy may be sufficiently large to contribute to the energy balance, but cannot dominate it, {and moreover is unlikely to be providing support against gravitational collapse on scales larger than the thermal Jeans length, ${\lambda_{J}\sim 0.03}$\,pc.}

We estimated the mass-to-flux ratio of OMC~1 to be {${\lambda_{obs}\sim\mphi}$}, less than but similar to unity, suggesting that the OMC~1 region is near magnetic criticality or slightly magnetically sub-critical.  {We also demonstrated that, in the absence of a magnetic field, the filament would be globally gravitationally unstable according to the \citet{ostriker1964} criterion.  However, the line mass of the filament is comparable to the magnetic critical line mass, suggesting that the filament is in or near magnetically-supported equilibrium.}

We determined the Alfv\'{e}n velocity in OMC~1 to be {${\alfven\pm\alfvenerr}$}\,km\,s$^{-1}$, and hence that the outflow could only produce Alfv\'{e}nic distortions on size scales of the order $10^{-3}$\,pc in 500 years (the approximate lifetime of the outflow), significantly smaller than the $\sim 10^{-1}$\,pc size scale of the hour-glass morphology.  We found that the typical velocity of the ballistic ejecta is significantly greater than the Alfv\'{e}n velocity, suggesting that perturbation of the field by the outflow would occur non-Alfv\'{e}nically.  However, the distance travelled by outflow ejecta is $\sim 10^{-2}$\,pc, smaller than the size scale of the hour-glass morphology.  Moreover, the energy density of the ballistic ejecta is several orders of magnitude smaller than the energy density of the magnetic field.  Hence, we concluded that the outflow is too young to have caused the large-scale morphology of the magnetic field in OMC~1.

We futher hypothesized that the direction of propagation and opening angle of the large-scale, ballistically ejected, BN/KL outflow (the `bullets of Orion') is constrained by the magnetic field geometry of the OMC~1 region.

We concluded that the gravitational {interactions} in OMC~1 {have} sufficient energy to be in or near equipartition with the magnetic field.  We hypothesized that the magnetic field morphology is the result of compression of an initially uniform and cylindrically-symmetric magnetic field by {some combination of the gravitational fragmentation of the filament and the gravitational} interaction of the BN/KL and S clumps.  {We further hypothesized that the magnetic field, while initially insufficiently strong to prevent the motion of material along the filament, may now have been increased by its compression to be sufficiently strong to slow or halt the flow of gas along the filament, and hence the interaction of the BN/KL and S clumps.  The hour-glass magnetic field may produce a cushioning and/or anchoring effect on the gas in the filament, causing any further interaction of the two clumps to be secular and mediated by ambipolar diffusion.}

\section{Acknowledgements}

The James Clerk Maxwell Telescope is operated by the East Asian Observatory on behalf of The National Astronomical Observatory of Japan, Academia Sinica Institute of Astronomy and Astrophysics, the Korea Astronomy and Space Science Institute, the National Astronomical Observatories of China and the Chinese Academy of Sciences (Grant No. XDB09000000), with additional funding support from the Science and Technology Facilities Council of the United Kingdom and participating universities in the United Kingdom and Canada.  The James Clerk Maxwell Telescope has historically been operated by the Joint Astronomy Centre on behalf of the Science and Technology Facilities Council of the United Kingdom, the National Research Council of Canada and the Netherlands Organisation for Scientific Research.  Additional funds for the construction of SCUBA-2 and POL-2 were provided by the Canada Foundation for Innovation.  The data used in this paper were taken under project codes M16AL004, M15BEC02 and MJLSG32.  KP and DWT would like to acknowledge support from the Science and Technology Facilities Council (STFC) under grant numbers ST/K002023/1 and ST/M000877/1 while this research was carried out.  Partial salary support for AP was provided by a Canadian Institute for Theoretical Astrophysics (CITA) National Fellowship.  CWL and WK were supported by Basic Science Research Program through the National Research Foundation of Korea (NRF) funded by the Ministry of Education, Science and Technology (CWL: NRF-2016R1A2B4012593) and the Ministry of Science, ICT \& Future Planning (WK: NRF-2016R1C1B2013642).  JCM acknowledges support from the European Research Council under the European Community's Horizon 2020 framework program (2014-2020) via the ERC Consolidator grant `From Cloud to Star Formation (CSF)' (project number 648505).  The Starlink software \citep{currie2014} is supported by the East Asian Observatory.  This research used the services of the Canadian Advanced Network for Astronomy Research (CANFAR) which in turn is supported by CANARIE, Compute Canada, University of Victoria, the National Research Council of Canada, and the Canadian Space Agency.  This research used the facilities of the Canadian Astronomy Data Centre operated by the National Research Council of Canada with the support of the Canadian Space Agency.  This research has made use of the NASA Astrophysics Data System. The authors wish to recognize and acknowledge the very significant cultural role and reverence that the summit of Mauna Kea has always had within the indigenous Hawaiian community. We are most fortunate to have the opportunity to conduct observations from this mountain.

\textit{Facilities:} James Clerk Maxwell Telescope (JCMT)
\textit{Software:} Starlink \citep{currie2014}, \textsc{smurf} (\citealt{berry2005}; \citealt{chapin2013}), Interactive Data Language (IDL)


\bibliographystyle{aasjournal}

\clearpage

\setcounter{figure}{0}
\setcounter{table}{0}
\renewcommand{\thefigure}{A\arabic{figure}}
\renewcommand{\thetable}{A\arabic{table}}

\section*{Appendix A: Effect of measurement uncertainties on angular dispersion - the general case}

 We here investigate the effect of the measurement uncertainty on recovered angular dispersion for the general case of a set of angles normally distributed about a mean of zero.  The aim of this investigation is to understand what, if any, systematic bias is introduced into a Chandrasekhar-Fermi analysis by measurement uncertainty on position angle.  We consider the generalised case in this appendix.  In Appendix B we discuss the effect of measurement uncertainty on the `unsharp masking' method used in this work.

We performed Monte Carlo simulations in order to determine the effect of measurement uncertainty {${\delta\theta}$} on the measured dispersion of angles $\sigma_{\theta}$ around the mean field direction.  {We generated 1000-element sets of distributions of angles ${\Delta\theta_{true}}$.  Each set ${\Delta\theta_{true}}$ was normally distributed about a mean of 0, with a specified standard deviation $\sigma_{\theta,true}$.}  {For each 1000-element set we generated an} accompanying set of measurement uncertainties $\delta\theta$, which were randomly drawn from {a uniform distribution with a range} $-30<\delta\theta\leq30$ degrees.

We note that the distribution of measurement uncertainties chosen for this analysis is not intended to be representative of the measurement uncertainties in our own data, which are ${ < 6}$ degrees everywhere where ${P/\delta P > 5}$.  We simply aim to understand the systematic effect of measurement error on recovered angular dispersion.

We note that there are two effects that deviate a measured magnetic field angle from the mean field direction: the physical dispersion in angle due to turbulence, and measurement error on position angle.  These two effects \emph{do not} add in quadrature: the former is a physical quantity which we aim to measure, while the latter is a statistical uncertainty on the former which is equally likely to move the measured deviation toward the mean field direction as away from it.  Thus, the observed deviation in angle, ${\Delta\theta}$, is given, for the ${i^{\rm th}}$ member of the distribution, by:
\begin{equation}
\Delta\theta_{i} = \Delta\theta_{true,i}+\delta\theta_{i}.
\end{equation}
Note that ${\delta\theta_{i}}$ can be negative.  The distribution ${\Delta\theta}$ thus contains both intrinsic dispersion and statistical uncertainties.  The observed dispersion in the data, ${\sigma_{\theta}}$, is then taken to be the standard deviation of ${\Delta\theta}$.

We repeated the process described above 10000 times for each of a set of specified values of ${\sigma_{\theta,true}}$.  The mean observed dispersion ${\bar{\sigma}_{\theta}}$ over those 10000 repeats was taken to be the best estimate of recovered dispersion, while the standard deviation of the observed dispersion, $\delta\bar{\sigma}_{\theta}$, was taken to represent 1-${\sigma}$ uncertainty on ${\bar{\sigma}_{\theta}}$.

\begin{figure}
  \centering
  \includegraphics[width=0.47\textwidth]{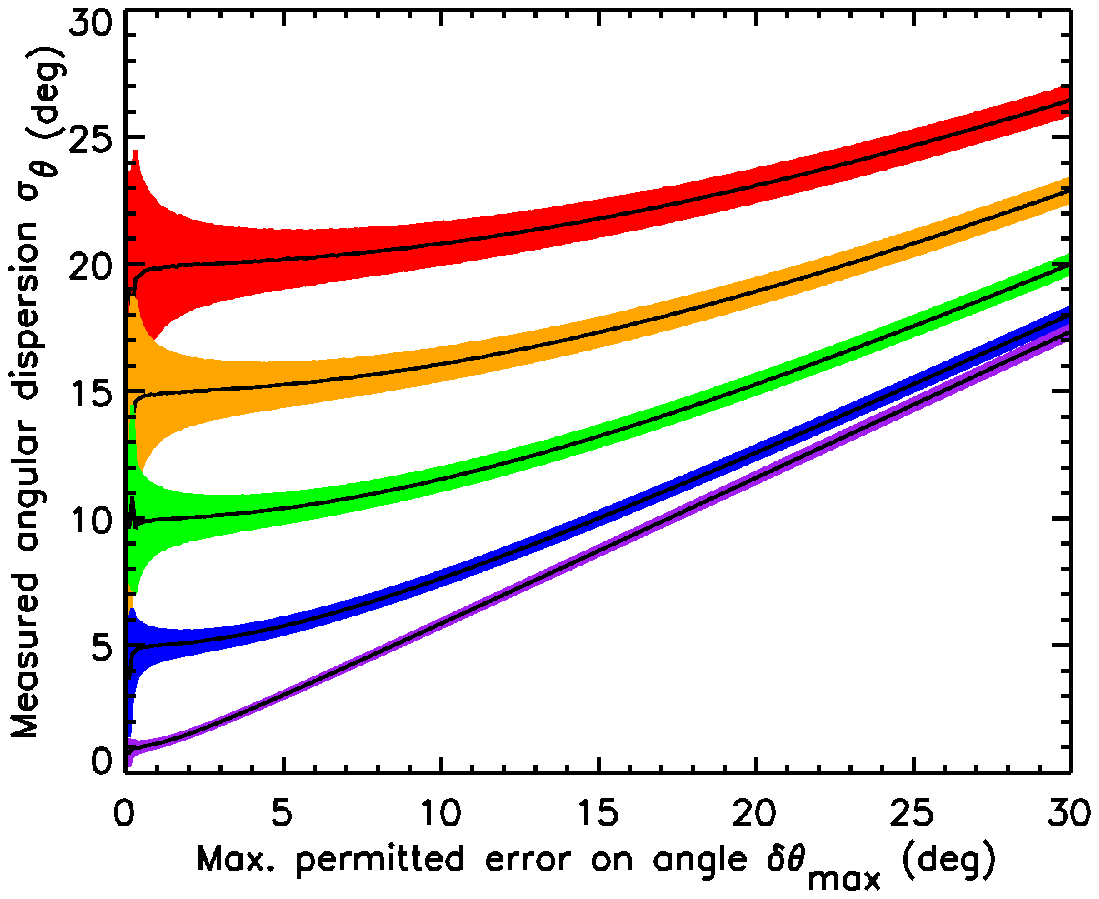}
  \caption{The behaviour of measured dispersion in angle as a function of maximum allowed uncertainty on angle, for underlying Gaussian distributions with widths 20 degrees (red), 15 degrees (orange), 10 degrees (green), 5 degrees (blue) and 1 degree (purple).  Angular uncertainties are drawn from a uniform distribution between 0 and 30 degrees.  Shaded regions indicate 1-$\sigma$ uncertainty on mean measured deviation.}
  \label{fig:figa4}
\end{figure}

The results of these Monte Carlo simulations are shown in Figure~\ref{fig:figa4}, {in which ${\bar{\sigma}_{\theta}}$ is plotted as a cumulative function of measurement uncertainty}.  We find two regimes of behaviour for {${\sigma_{\theta}}$}.  The measured angular dispersion {${\sigma_{\theta}}$} agrees well with the true angular dispersion {${\sigma_{\theta,true}}$} in well-characterized pixels, when the maximum uncertainty on angle $\delta\theta_{max}$ is smaller than the true dispersion of the distribution of angles, i.e. $\sigma_{\theta}\approx\sigma_{\theta,true}$ when $\delta\theta_{max}\ll\sigma_{\theta,true}$.  However, when {${\delta\theta_{max}}$} is equal to or greater than {${\sigma_{\theta,true}}$}, {${\sigma_{\theta}}$} increases approximately linearly with {${\delta\theta_{max}}$}, i.e. $\sigma_{\theta}>\sigma_{\theta,true}$ when $\delta\theta_{max}\gtrsim\sigma_{\theta,true}$.

Figure~\ref{fig:figa4} shows that when $\delta\theta_{max}$ is equal to or greater than $\sigma_{\theta,true}$ (i.e. when the data are less well-characterized than our POL-2 data, as discussed below), then the systematic effect of the angular uncertainty on $\sigma_{\theta}$ must be accounted for.  Previous attempts to do this have included subtraction of mean measurement uncertainty in quadrature (e.g \citealt{crutcher2004}, observing with SCUPOL, with measurement uncertainties expected to be sufficiently large to be in the regime in which correction was required).

\clearpage

\setcounter{figure}{0}
\setcounter{table}{0}
\renewcommand{\thefigure}{B\arabic{figure}}
\renewcommand{\thetable}{B\arabic{table}}

\section*{Appendix B: Demonstration of the `unsharp masking' method of determining angular dispersion}

We tested our `unsharp masking' method of determining angular dispersion by applying it to a series of sets of synthetic observations.  Our method was as follows: (1) we generated a set of parabolas with a specified focal length, (2) we applied a Gaussian angular dispersion to these data, the standard deviation of which is the measurement that we wish to recover, (3) we applied a set of measurement errors to the data, drawn from a Gaussian distribution with a specified standard deviation, (4) we smoothed the `dispersion + errors' map with a ${3\times 3}$ boxcar filter, (5) we subtracted the smoothed map from the `dispersion + errors' map, (6) we measured the standard deviation in the residuals map, and compared it to the input standard deviation.  We note that we implicitly assume throughout this analysis that the pixels over which we smooth are linearly independent, i.e., for real data, pixel size $\gtrsim$ beam size.

We tested this method for a range of parabola focal lengths.  Our input parabolas took the form
\begin{equation}
y_{h} = \frac{1}{4f}x^{2} + h
\end{equation}
where $h$ is an integer offset.  We specified the focal length $f$ in terms of its ratio to the size of the smoothing box size ${S}$, in this case 3 pixels.

\begin{figure}
\centering
\includegraphics[width=0.4\textwidth]{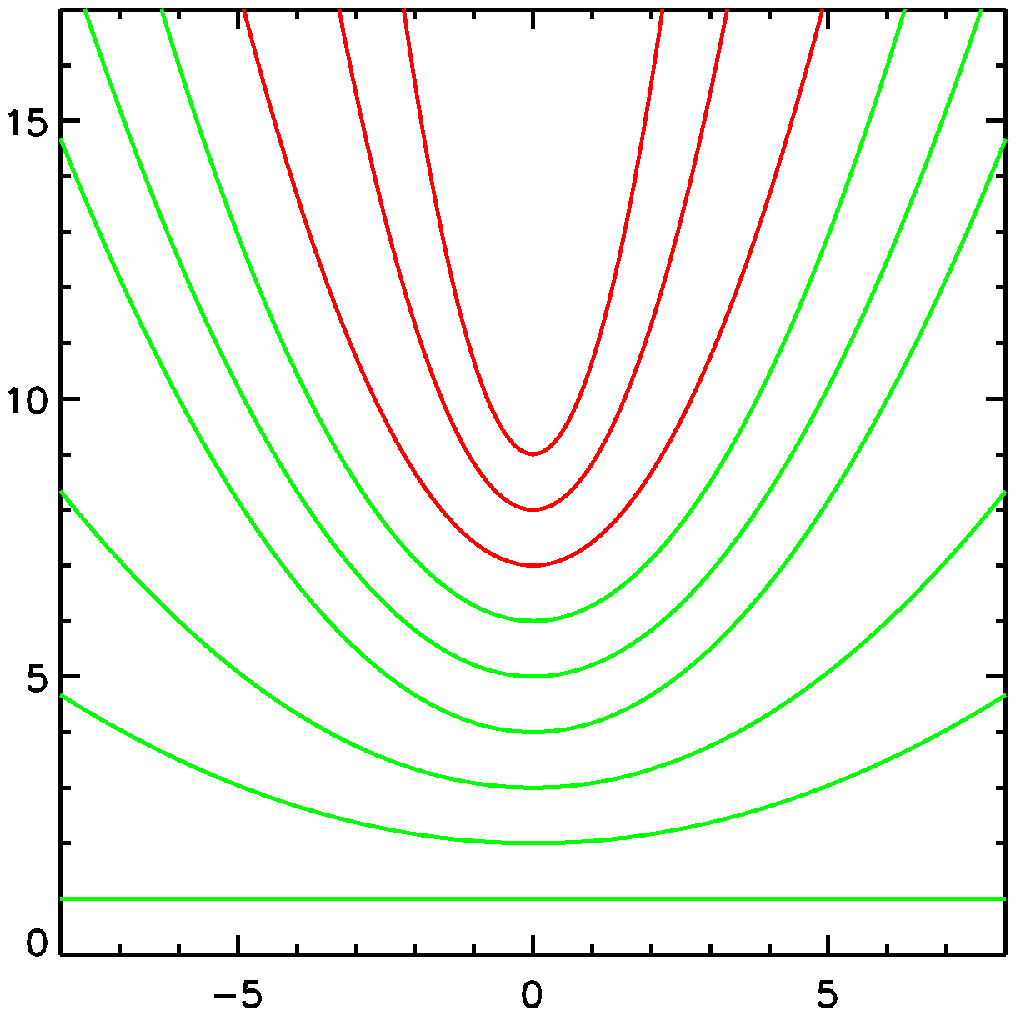}
\caption{The model parabolas which were tested, with focal lengths $f$ of $\infty$, 6.0, 3.0, 1.5, 1.2, 0.9, 0.6, 0.3 and 0.15 pixels (bottom to top; $y$ offset is arbitrary).  Parabolas shown in green were sufficiently shallow for angular dispersion to be accurately recovered (i.e. ${\sigma_{\theta}\approx\sigma_{\theta,true}}$); the method fails for parabolas shown in red.}
\label{fig:parabolas}
\end{figure}

The set of parabolas that we tested are shown in Figure~\ref{fig:parabolas}.  We modelled parabolas with ${f/S}$ values of 2.0, 1.0, 0.5, 0.4, 0.3, 0.2, 0.1 and 0.05; i.e. with focal lengths $f$ of 6.0, 3.0, 1.5, 1.2, 0.9, 0.6, 0.3 and 0.15 pixels.  We also tested the zero-curvature case, i.e. ${f=\infty}$.

{We modelled angular dispersion ${\sigma_{\theta,true}}$ in the range ${1\leq\sigma_{\theta,true}\leq10}$ degrees.  We drew uncertainties ${\delta\theta}$ from uniform distributions with ranges ${-\delta\theta_{max}\leq\delta\theta\leq\delta\theta_{max}}$, choosing values of ${\delta\theta_{max}}$ in the range ${0-10}$ degrees.  In any given pixel with coordinates ${(i,j)}$, the measured angle ${\theta_{obs}}$ is given by}
\begin{equation}
{\theta_{obs,i,j} = \tan^{-1}\left(\frac{i}{2f}\right) + (\Delta\theta)_{true,i,j} + \delta\theta_{obs,i,j},}
\end{equation}
{where the intrinsic deviation in the field direction ${(\Delta\theta)_{true}}$ is drawn from a probability distribution with Gaussian widths ${\sigma_{\theta,true}}$, and the uncertainty on measurement angle is drawn from a uniform distribution specified by the value of ${\delta\theta_{max}}$.}

{The mean field direction ${\langle\theta\rangle}$ is given, in a pixel with coordinates ${(i,j)}$, by}
\begin{equation}
{\langle\theta\rangle_{i,j} = \frac{1}{S^{2}}\sum_{l=-\frac{S-1}{2}}^{\frac{S-1}{2}}\sum_{k=-\frac{S-1}{2}}^{\frac{S-1}{2}}\theta_{obs,i+k,j+l}}
\end{equation}
{where box size $S$ is an odd integer greater than 1.  The recovered deviation in mean field direction $\Delta\theta$ in pixel $(i,j)$ is then given by }
\begin{equation}
{\Delta\theta_{i,j} = \theta_{obs,i,j} - \langle\theta\rangle_{i,j}.}
\end{equation}
{The recovered dispersion ${\sigma_{\theta}}$ is then the standard deviation of ${\Delta\theta}$.}

\begin{figure}
\centering
\includegraphics[width=\textwidth]{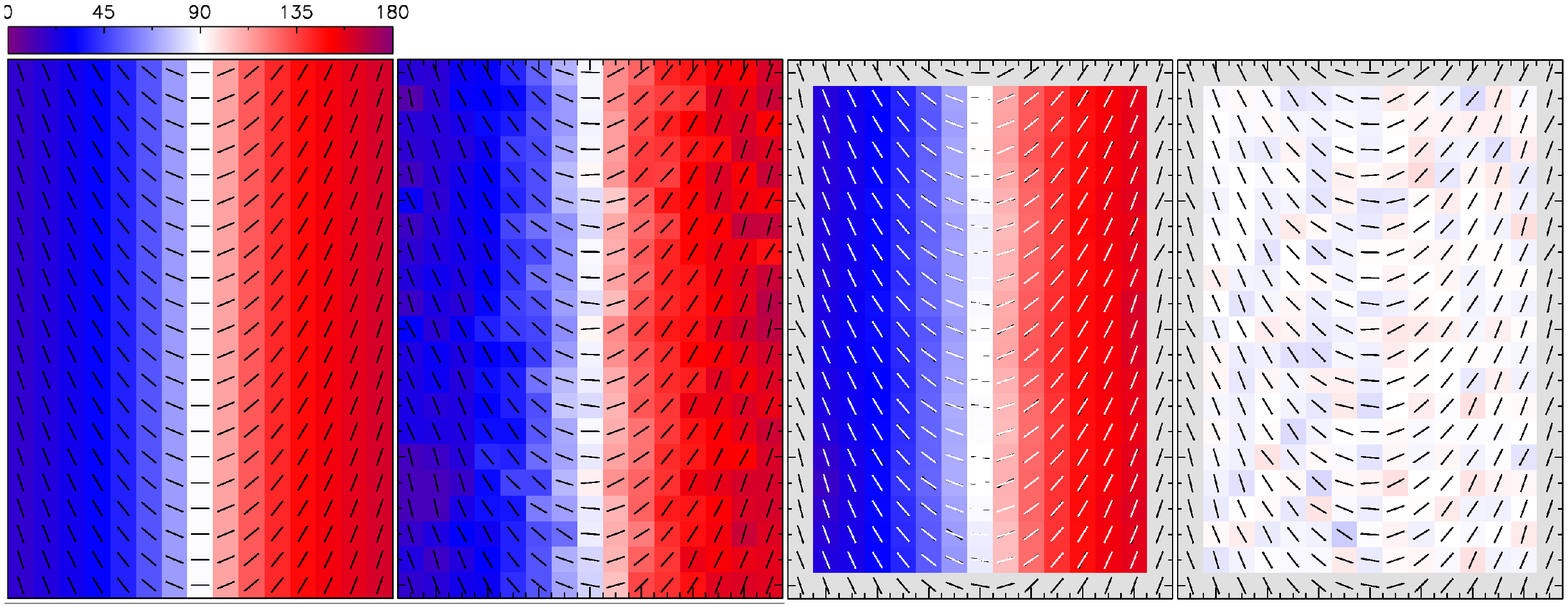}
\caption{(a) Underlying field model; ${f/S = 0.4}$.  (b) Model observations, with ${\sigma_{\theta,true} = 4.0}$ degrees and ${\delta\theta_{max}<2}$ degrees. (c) Smoothed map, showing estimated mean field direction ${\langle\theta\rangle}$.  (d) Residuals, ${\Delta\theta = \theta_{obs} - \langle\theta\rangle}$.  Black vectors indicate magnetic field direction.  In panel (c), white vectors indicate smoothed magnetic field direction, for comparison.}
\label{fig:model_obs}
\end{figure}

{We estimated the mean value of ${\sigma_{\theta}}$ and the standard deviation on that value by performing Monte Carlo simulations.  For each of our chosen values of ${f/S}$, ${\sigma_{\theta,true}}$ and ${\delta\theta_{max}}$, we drew 500 sets of angular deviations and uncertainties, and found the mean and standard deviation of the values of ${\sigma_{\theta}}$ recovered from these 500 data sets.  An example model is shown in Figure~\ref{fig:model_obs}, with ${f/S = 0.4}$, ${\sigma_{\theta,true} = 4.0}$ degrees and ${\delta\theta_{max}<2}$ degrees, similar to what we measure in high-signal-to-noise regions of OMC~1.}

\begin{figure}
\centering
\includegraphics[width=0.5\textwidth]{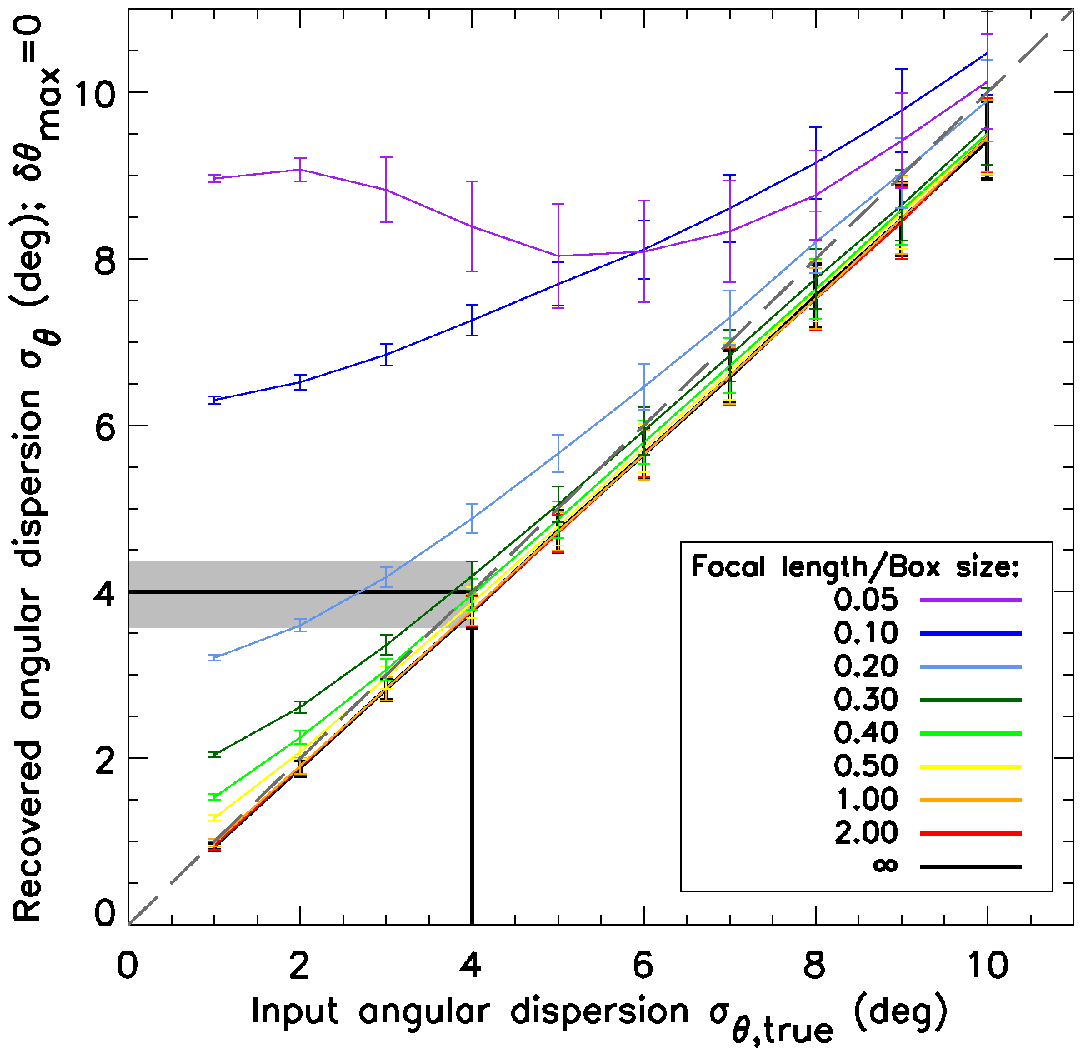}
\caption{{Comparison of input and recovered angular dispersions in the limiting case where measurement errors are negligible (note that the ${f/S=2.0}$ case tends closely to ${f/S=\infty}$).  Dashed line shows the 1:1 line.  Solid black line marks ${\sigma_{\theta,true}=4.0}$ degrees; the grey shaded region shows the range of ${\sigma_{\theta}}$ values recovered for field curvatures seen in OMC~1 when ${\sigma_{\theta,true}=4.0}$ degrees.}}
\label{fig:model_results}
\end{figure}

{The results of our synthetic observations in the limiting case where measurement errors are negligible are shown in Figure~\ref{fig:model_results}.}

\subsection*{Effect of field curvature}

Our synthetic observations predict that for field curvatures ${f/S \gtrsim 0.3}$, this method recovers ${\sigma_{\theta,true}}$  with reasonable accuracy (see Figure~\ref{fig:model_results}).  While ${f/S \gtrsim 0.3}$, ${\sigma_{\theta,true}}$ is recovered well while ${\sigma_{\theta,true}\gtrsim 3.0}$ degrees.

This method overestimates ${\sigma_{\theta,true}}$ for very low values of ${\sigma_{\theta,true}}$, as can be seen in Figure~\ref{fig:model_results}.  The method fails for very high degrees of curvature in the underlying field ($f/S \ll 1$), as best seen in the ${f/S = 0.10}$ and ${f/S=0.05}$ cases in Figure~\ref{fig:model_results}.  The failure of the method at low ${\sigma_{\theta,true}}$ and at high field curvature ${f/S}$ is likely to be due to the systematic variation in field direction over the box size (due to the changing direction of the underlying field) being comparable to or greater than the random variation in field direction due to the dispersion on position angle.  The difference between the mean angle and the observed angle (${\theta_{obs} - \langle\theta\rangle}$) thus becomes a measure of field curvature rather than of angular dispersion.

Degrees of field curvature ${f/S}$ for which ${\sigma_{\theta,true}}$ is recovered well (for a box size of of ${3\times3}$ pixels) are shown in green in Figure~\ref{fig:parabolas}, while field curvatures ${f/S}$ for which the method fails are shown in red.  We define failure in general as $\sigma_{\theta}>\sigma_{\theta,true}$ for all $\sigma_{\theta,true}$, and in our specific case as $\sigma_{\theta}$ differing from $\sigma_{\theta,true}$ by $\gtrsim 10$\% when $\sigma_{\theta,true}\sim 4$ degrees.  Comparison of this figure with the magnetic field in the high-S/N region of OMC~1 over which we perform the analysis (Figure~\ref{fig:angles}c) shows that over the vast majority of this region the field curvature is definitively in the regime in which angular dispersion is recovered well.  We estimate that in the highest-curvature region of OMC~1, the field curvature reaches a maximum of ${f\approx0.83}$, i.e. ${f/S\approx 0.28}$.  Hence, over all of the high-S/N region of OMC~1, the field curvature is in the regime ${f/S\lesssim 0.3}$, and so this method does not overestimate the angular dispersion in this case.  Another check on this is examination of the residual map (Figure~\ref{fig:angles}c); we do not see systematically higher residuals in regions of higher field curvature, which suggests that we are accurately recovering the true angular dispersion.

\subsection*{Effect of box size}

We note that, in the absence of significant effects from field curvature, there is a slight tendency for this method to underestimate larger values of ${\sigma_{\theta,true}}$.  In the limiting case of no curvature (${f/S=\infty}$), this method systematically recovers a value of ${\sigma_{\theta}}$ which is ${0.94\times}$ the true angular dispersion ${\sigma_{\theta,true}}$, and recovered angular dispersions ${\sigma_{\theta}}$ for fields which have curvature tend to this value as ${\sigma_{\theta,true}}$ becomes large (see Figure~\ref{fig:model_results}).

This systematic underestimation is a result of the ${3\times3}$-pixel boxcar filter not being sufficiently large to sample the full range of variation in angle across the map, and so underestimating ${\sigma_{\theta,true}}$.  In the zero-curvature case (${f/S=\infty}$), we find that as smoothing box size ${S}$ increases, ${\sigma_{\theta}}$ tends toward ${\sigma_{\theta,true}}$: a ${5\times 5}$-pixel box recovers ${0.981\times\sigma_{\theta,true}}$, while a ${7\times 7}$-pixel box recovers ${0.984\times\sigma_{\theta,true}}$.  However, as ${S}$ is increased, the effect of systematic variation in field direction over the box becomes more significant, and for even relatively shallow field curvatures the angular dispersion is significantly overestimated as a result.

The effect of this systematic underestimation is minimal while ${\sigma_{\theta,true}}$ is small; moreover, for small ${\sigma_{\theta,true}}$ the slight overestimation of ${\sigma_{\theta}}$ due to systematic variation in the field direction over the smoothing box mitigates against this effect (see Figure~\ref{fig:model_results}).  For relatively large angular dispersions (${\sigma_{\theta,true}\gtrsim 8}$ degrees; considerably higher than is seen in OMC~1), the systematic underestimation produces a ${\gtrsim 1-\sigma}$ offset between  ${\sigma_{\theta}}$ and ${\sigma_{\theta,true}}$.  If a data set were to fall into that regime, ${\sigma_{\theta}}$ could be corrected for this systematic effect simply by multiplying it by a factor 1.06.

For our data, the dispersion that we measure, ${\langle\sigma_{\theta}\rangle = \angdisp\pm\angdisperr}$ degrees, falls into the regime where any systematic effects from field curvature and box size mitigate against each other; Figure~\ref{fig:model_results} shows that values of ${\sigma_{\theta,true}}$ in the range ${3 - 8}$ degrees will be recovered accurately for a wide range of field curvatures ${f/S}$.  We note also that the systematic offsets in angular dispersion which are predicted by our synthetic observations are smaller than the statistical uncertainty on our result.

\subsection*{Effect of measurement errors}

\begin{figure}
\centering
\includegraphics[width=0.4\textwidth]{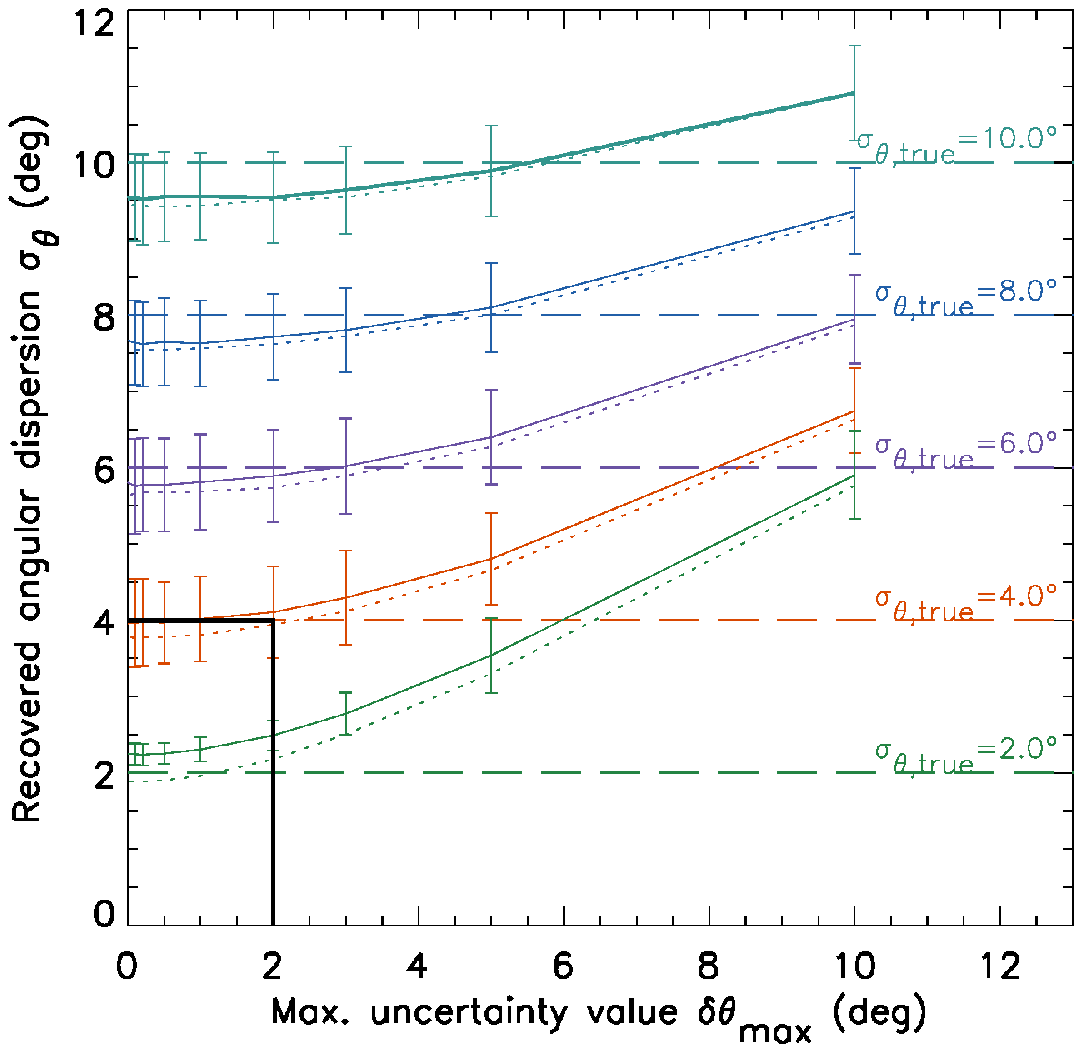}
\caption{{The effect of measurement errors on recovered angular dispersion ${\sigma_{\theta}}$, for a field curvature ${f = 1.2}$ (${f/S = 0.4}$; solid lines), and for the zero-curvature case (${f/S=\infty}$; dashed lines).  Solid black line shows ${\sigma_{\theta}}$ when ${\sigma_{\theta,true}=4.0}$ degrees and ${\delta\theta_{max}<2.0}$ degrees, as is the case for the well-characterised pixels in OMC~1.}}
\label{fig:model_errors}
\end{figure}

{As demonstrated in Appendix~A, over well-characterised pixels, and in the absence of field curvature, ${\sigma_{\theta}}$ should not be systematically altered by measurement errors.  We here test whether this result holds when field curvature is included.   We drew uncertainties ${\delta\theta}$ from uniform distributions with ranges ${-\delta\theta_{max}\leq\delta\theta\leq\delta\theta_{max}}$, choosing values of ${\delta\theta_{max}}$ in the range ${0-10}$ degrees.  The results of these tests, for a field curvature ${f/S = 0.4}$, are shown as solid lines in Figure~\ref{fig:model_errors}.  The results of these tests for the zero-curvature unsharp-masking case are shown as dotted lines on the same figure.}

{We see that as in the generalised zero-curvature case, ${\sigma_{\theta}}$ is not altered by measurement errors while those measurement errors are small.  However, as previously, ${\sigma_{\theta}}$ increases approximately linearly with ${\delta\theta_{max}}$ when ${\delta\theta_{max}\gtrsim\sigma_{\theta,true}}$.  This is true both for the zero-curvature and ${f/S=0.4}$ unsharp-masking examples shown in Figure~\ref{fig:model_errors}.  We do not see any tendency for ${\sigma_{\theta}}$ to deviate more rapidly from ${\sigma_{\theta,true}}$ as ${\delta\theta_{max}}$ increases in the ${f/S=0.4}$ case than in the zero-curvature case.}

{We find that for ${\sigma_{\theta,true}\approx 4}$ degrees, the systematic effect of measurement error on ${\sigma_{\theta}}$ is minimal while ${\delta\theta_{max}\lesssim 2}$ degrees, as is the case in the region of OMC~1 over which we perform our analysis (discussed below).}

\subsection*{Application to OMC~1 data}

As discussed above, if ${\delta\theta_{max}\lesssim 2}$, then ${\sigma_{\theta}\approx\sigma_{\theta,true}}$ (if ${f/S}$ is sufficiently small).  We thus restrict our application of the unsharp-masking method in OMC~1 to those pixels for which the maximum uncertainty in any pixel included in the smoothing box is ${< 2.0}$ degrees.  Uncertainties on position angle are calculated by \textit{pol2stack} from the variances on the ${Q}$ and ${U}$ values in each pixel in the coadded ${Q}$ and ${U}$ maps from which the vector properties are calculated, using standard error propagation (see Section~\ref{sec:observations}).

\begin{figure}
  \centering
  \includegraphics[width=0.47\textwidth]{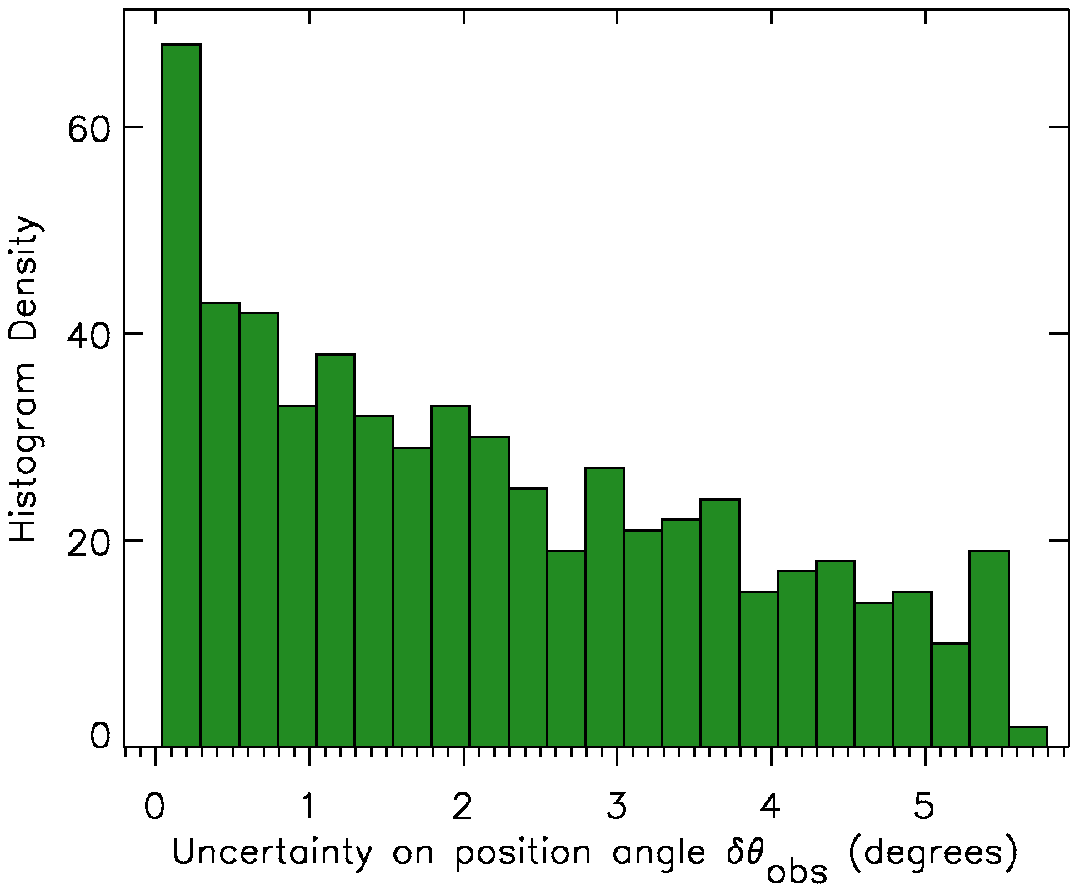}
  \caption{The distribution in uncertainties on measured position angles $\delta\theta_{obs}$ for valid pixels in OMC~1 with $(P/\delta P)\geq5$.}
  \label{fig:figa1}
\end{figure}

Figure \ref{fig:figa1} shows the distribution of uncertainties $\delta\theta_{obs}$ on the pixels in OMC~1 for which $(P/\delta P)\geq5$ and which are not excluded from the analysis for containing changes in angle $\geq 90$ degrees in their smoothing box (see Section~\ref{sec:angdisp}).

The pixels in OMC~1 with low measurement uncertainties can be seen as a largely contiguous region with low residuals in Figure~\ref{fig:angles}c.  {{Figure~\ref{fig:figa6} shows the variation in ${\delta\theta_{max}}$ across OMC~1, with the region over which ${\delta\theta_{max}<2.0}$ degrees outlined in black.}  This contiguous region includes the high-density region of OMC~1: the BN/KL and S regions, the region between them, and most of the region in which the magnetic field shows an hour-glass morphology.

\begin{figure}
  \centering
  \includegraphics[width=0.4\textwidth]{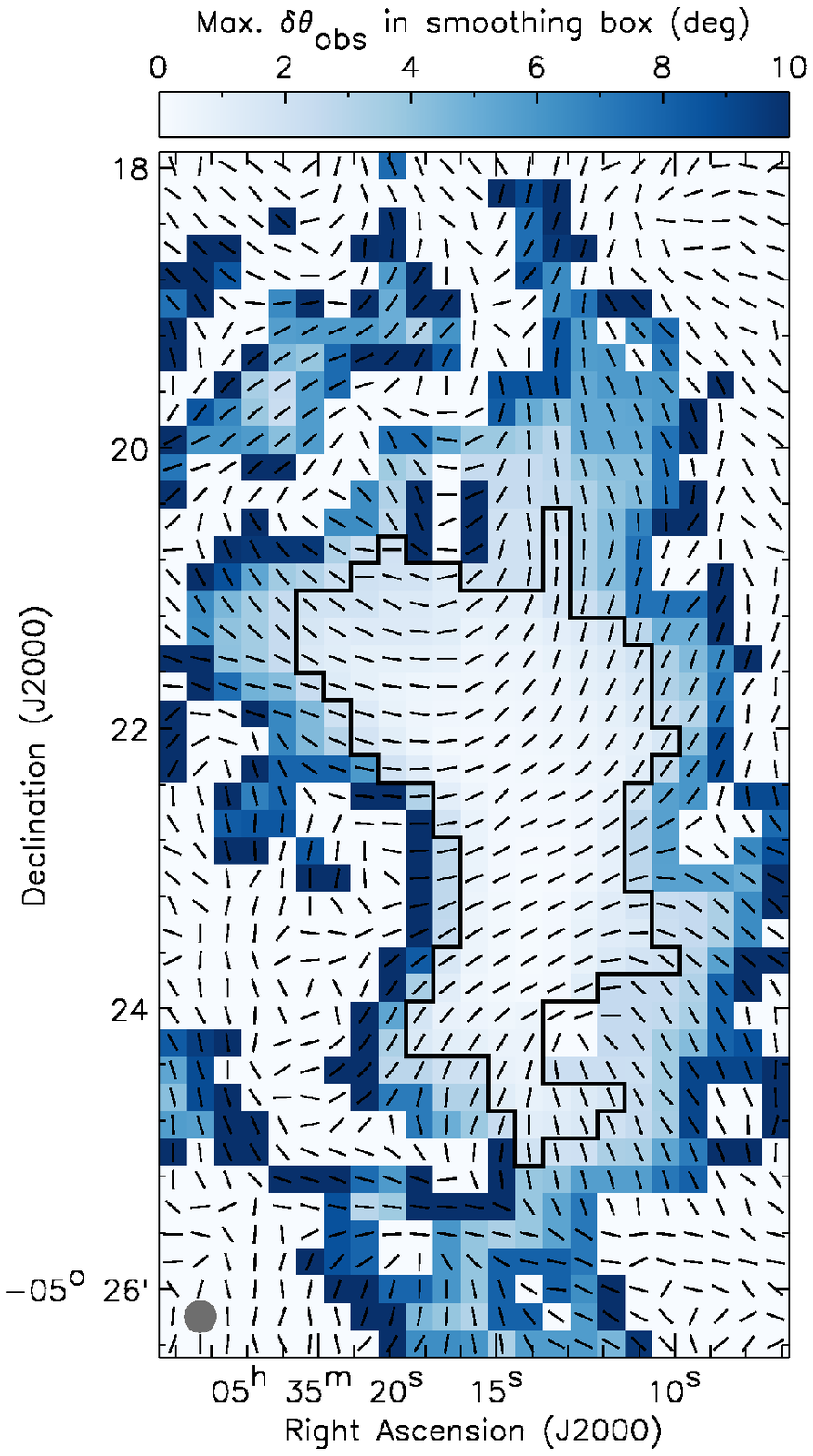}
  \caption{{Map of maximum uncertainty ${\delta\theta_{max}}$ within the ${3\times3}$ smoothing box associated with each pixel across OMC~1.  Region outlined in black indicates ${\delta\theta_{max}<2.0}$.  Grey pixels have either ${P/DP < 5}$ or changes of angle ${> 90}$ degrees in their smoothing box.}}
  \label{fig:figa6}
\end{figure}

{We took the standard deviation of the cumulative distribution of ${\Delta\theta}$ as a function of increasing ${\delta\theta_{max}}$ in order to determine a representative value of ${\sigma_{\theta}}$ for OMC~1.  This is plotted in Figure~\ref{fig:ang_dev}.  Taking the mean of the standard deviations of the distributions containing only the best-characterized pixels (${\displo<\delta\theta_{max}<\disphi}$ degrees; up to \incpix\ pixels), we found a dispersion of ${\langle\sigma_{\theta}\rangle = \angdisp\pm \angdisperr}$ degrees.  We thus adopt this value of angular dispersion for our Chandrasekhar-Fermi analysis of Orion A, as being determined from the best-characterized pixels, and from the area of most relevance for our scientific analysis.  As shown by Figures~\ref{fig:model_results} and \ref{fig:model_errors}, the angular dispersion which we determine from these pixels will be an accurate estimate of the true angular dispersion in OMC~1.}

\label{lastpage}

\end{document}